\documentclass[%
preprint,
 amsmath,amssymb,
 aps,
floatfix
]{revtex4-1}

\usepackage{color}
\usepackage{graphicx}
\usepackage{dcolumn}
\usepackage{bm}
\usepackage{ulem}
\usepackage{braket}
\usepackage{enumerate}
\usepackage{rotating}
\usepackage{comment}
\usepackage{tikz}
\usepackage{pgfplots}

\begin{document}

\title{Natural orbitals in multiconfiguration calculations of hyperfine structure parameters}

\author{Sacha Schiffmann}
\affiliation{Spectroscopy, Quantum Chemistry and Atmospheric Remote Sensing  \\ Universit\'e libre de Bruxelles,  CP160/09, 1050 Brussels, Belgium}
\affiliation{Division of Mathematical Physics, Department of Physics, Lund University, SE-22100 Lund, Sweden}

\author{Michel Godefroid}
 \email{mrgodef@ulb.ac.be}
\affiliation{Spectroscopy, Quantum Chemistry and Atmospheric Remote Sensing  \\ Universit\'e libre de Bruxelles,  CP160/09, 1050 Brussels, Belgium}
\author{J\"orgen Ekman}
\affiliation{Department of Materials Science and Applied Mathematics, Malm\"o University, SE-20506 Malm\"o, Sweden}
\author{Per J\"onsson}
\affiliation{Department of Materials Science and Applied Mathematics, Malm\"o University, SE-20506 Malm\"o, Sweden}
\author{Charlotte Froese Fischer}
\affiliation{Department of Computer Science, University of British Columbia, 2366 Main Mall,
Vancouver, BC V6T1Z4, Canada}

\date{\today}

\begin{abstract}
We are reinvestigating the hyperfine structure of sodium using a fully relativistic multiconfiguration approach. In the fully relativistic approach, the computational strategy somewhat differs from  the original nonrelativistic counterpart used in~\cite{Jonetal:96b}. Numerical instabilities force us to use a layer-by-layer approach that has some broad  unexpected effects. Core correlation is found to be significant and therefore requires to be described in an adequate orbital basis. The natural-orbital basis provides an interesting alternative to the orbital basis from the layer-by-layer approach, allowing us to overcome some deficits of the latter, giving rise to magnetic dipole hyperfine structure constant values in excellent agreement with observations. Effort is made to assess the reliability of the natural-orbital bases and to illustrate their efficiency.
\end{abstract}

\maketitle
\section{Introduction}
\label{sec:NR_hfs}
The hyperfine structure of neutral sodium was investigated by J\"onsson et al.~\cite{Jonetal:96b} using the non-relativistic multiconfiguration Hartree-Fock (MCHF) approach~\cite{Froetal:97b}. Large-scale calculations were performed to estimate not only the hyperfine structure constants of the [Ne]$3s \ ^{2}\!S$ and [Ne]$3p \ ^{2}\!P^\circ$ terms, but also the transition probability of the $3s \ ^{2}\!S - 3p \ ^{2}\!P^\circ$ resonance transition. Their calculations resolved a long-standing discrepancy between  theory and experiment of the transition probability and provided accurate values for the hyperfine structure constants. Their optimization strategy consisted of the simultaneous variation of all correlation orbitals together with the spectroscopic $3s$/$3p$ orbital for the $3s \ ^{2}\!S$/$3p \ ^{2}\!P^\circ$ state. The simultaneous optimization of all correlation orbitals together with the spectroscopic valence orbitals, that we will refer to as the ``full variational'' (FV) approach, has been widely used in non-relativistic calculations~\cite{Bieetal:2018a,CarGod:2013a,Caretal:2010a}. Due to numerical convergence issues, the relativistic counterpart of the MCHF method, the multiconfiguration Dirac-Hartree-Fock (MCDHF) method, employs almost exclusively a layer-by-layer (LBL) strategy~\cite{FroGod:2019a,Papetal:2019a} in which only the newly introduced orbitals for the layer considered are optimized while the remaining ones are kept frozen. In this context, a layer is a set of new orbitals to be optimized including one orbital per angular momentum symmetry. 
To first-order, the total energy should converge towards the same limit in the two optimization schemes, if considering orbital active spaces that are large enough in the single- and double-excitation process.
The LBL approach is attractive as the computation time for each new layer is much shorter than the corresponding computation time of the FV approach. The price to pay for the LBL strategy is  a larger active set of correlation orbitals to compensate for the lost degrees of freedom in the variational process relatively to the FV method. \\

As a complement to~\cite{Jonetal:96b}, we performed non-relativistic calculations of the hyperfine structure constant $A_{1/2}$ of the sodium ground state using the Atomic Structure Package ({\sc ATSP})~\cite{Froetal:2007a} keeping the same correlation model than in~\cite{Jonetal:96b} for describing the core-valence electron correlation, but adopting the LBL instead of the FV approach. In agreement with its lower variational flexibility, the layer-by-layer approach requires  additional orbital layer(s) to reach convergence of the hyperfine constant and to reproduce the original result. To improve the correlation model, core-core (CC) correlation is added in the last step through configuration interaction (CI) calculations. The comparison of the final values reveals a surprisingly large difference ($\sim 23$ MHz) between the two approaches, suggesting that the magnetic dipole hyperfine constant is considerably underestimated in the LBL approach.

The MCDHF method as implemented in the GRASP2K and GRASP2018 packages~\cite{Jonetal:2013a, Fischer:2019aa}, often fails in optimizing all orbitals simultaneously, forcing the user to adopt the layer-by-layer optimization strategy. The {\it a priori} unexpected discrepancy between the LBL and FV $A_{1/2}$ values observed in the non-relativistic framework is therefore disturbing and raises the relevant question of reliability of the final LBL relativistic hyperfine structure value, the only one that can be estimated so far with the MCDHF codes. In the present paper, we investigate the use of different orbital bases in hyperfine structure calculations in both the non-relativistic and relativistic frameworks and evaluate the usefulness and reliability of natural orbitals (NO). \\

The relevant non-relativistic theoretical background can be found in~\cite{Jonetal:96b}. Sections~\ref{sec:MCDHF} and~\ref{sec:RHFS} briefly describe respectively the MCDHF and the hyperfine structure theories. Natural orbitals are introduced in Section~\ref{sec:NO}. Section~\ref{sec:NaGS} presents their application to the sodium ground state hyperfine structure while Section~\ref{sec:iso} extends the analysis to sodium-like ions ground state and sodium excited states.

\section{The MCDHF theory}
\label{sec:MCDHF}
In the multiconfiguration Dirac-Hartree-Fock (MCDHF) method~\cite{Gra:2006a}, an atomic state of total angular momentum $J$ and parity $\pi$ is expanded over configuration state functions (CSFs) as
\begin{equation}
    \label{eq:ASF}
    \Psi(\Gamma J\pi) = \sum_{j=1}^{N_{\text{CSF}}} c_{j} \Phi(\gamma_{j}J\pi) \ ,
\end{equation}
where $\gamma_{j}$ specifies the angular coupling tree of the $j^\text{th}$ CSF. CSFs are themselves built as sums of anti-symmetric products of one-electron Dirac spinors
\begin{equation}
\label{eq:Dirac_Spinor}
\phi_{n\kappa m}(\boldsymbol{r},\sigma) = \frac{1}{r}\begin{pmatrix} P_{n\kappa}(r)\chi_{\kappa m}(\theta,\phi) \\ \mbox{i}Q_{n\kappa}(r)\chi_{-\kappa m}(\theta,\phi) \end{pmatrix} \ ,
\end{equation}
which satisfy the following normalization condition $\int_{0}^{\infty} \left[(P_{n\kappa}(r))^{2}+ (Q_{n\kappa}(r))^{2}\right] dr = 1$.
The $P_{n\kappa}(r)$ and $Q_{n\kappa}(r)$ radial orbitals, defined on an exponential grid, are determined by solving iteratively coupled integro-differential equations derived by applying the variational principle to the energy functional based on the $N$-electron Dirac-Coulomb Hamiltonian
\begin{equation}
    H_{\text{DC}} = \sum_{j=1}^{N} \left[c\; \boldsymbol{\alpha_{j}}\cdot \boldsymbol{p} _{j}+ (\beta_{j} - 1)c^{2} + V_{nuc}(r_{j}) \right]+ \sum_{i>j=1}^{N}\frac{1}{r_{ji}} \ .
\end{equation}
The orthogonality constraints $\braket{\phi_{n\kappa m}|\phi_{n'\kappa m}}=\delta_{nn'}$ are introduced in the functional through appropriate Lagrange parameters~\cite{Froetal:2016a}. For a given set of one-electron orbitals, the $c_{j}$ mixing coefficients of Eq.~\ref{eq:ASF} are normalized solutions of the secular equations. The MCDHF method iteratively computes the radials orbitals and the mixing coefficients until self-consistency. \\
As a last step, the MCDHF method is usually followed by relativistic configuration interaction (CI) calculations to include higher-order excitations and/or additional interactions in the Hamiltonian such as the long-wavelength approximation Breit interaction
\begin{equation}
    H_{\text{Breit}} = - \sum_{i>j=1}^{N} \frac{1}{2r_{ji}} \left[\boldsymbol{\alpha_{j}} \cdot \boldsymbol{\alpha_{i}} + \frac{(\boldsymbol{\alpha_{j}} \cdot \boldsymbol{r_{ji}})(\boldsymbol{\alpha_{i}} \cdot \boldsymbol{r_{ji}})}{r_{ji}^{2}} \right] \ ,
\end{equation}
or QED corrections~\cite{Froetal:2016a,Aouetal:2018a}.

\section{Hyperfine structure theory}
\label{sec:RHFS}
The hyperfine structure results from the non central interaction between the electromagnetic multipole moments of the nucleus and the electron cloud. The hyperfine structure Hamiltonian is a sum over multipole moments
\begin{equation}
    \label{eq:H_hfs}
    H_{\text{hfs}} = \sum_{k=1}^{\infty} \boldsymbol{T}^{(k)} \cdot \boldsymbol{M}^{(k)}
\end{equation}
where $\boldsymbol{T}^{(k)}$ and $\boldsymbol{M}^{(k)}$ are electronic and nuclear tensorial operators of rank $k$, respectively. The coupling of the nuclear and atomic electromagnetic properties leads us to combine the total electronic angular moment $\textbf{J}$ and the nuclear spin $\textbf{I}$ in a total angular momentum $\textbf{F}=\textbf{J}+\textbf{I}$~\cite{Lindgren1974}. \\
In our calculations, the multipole expansion of Eq.~\ref{eq:H_hfs} is truncated to its two lowest rank terms, describing, respectively, the magnetic dipole interaction ($k=1$) and the electric quadrupole interaction ($k=2$). The corresponding electronic tensors  are defined as
\begin{equation}
    \label{eq:T1}
    \boldsymbol{T}^{(1)} = -\mbox{i} \alpha \sum_{i=1}^{N}(\boldsymbol{\alpha}_{i}\cdot \boldsymbol{l}_{i} \boldsymbol{C}^{(1)}(i))r_{i}^{-2}
\end{equation}
and
\begin{equation}
    \label{eq:T2}
    \boldsymbol{T}^{(2)} = - \sum_{i=1}^{N} \boldsymbol{C}^{(2)}(i)r_{i}^{-3} \ .
\end{equation}
The energy splitting induced by the hyperfine interaction is commonly expressed in term of the hyperfine magnetic dipole constant $A_{J}$ and the hyperfine electric quadrupole constant $B_{J}$~\cite{Froetal:97b}, which are defined by the relations
\begin{equation}
    A_{J} = \frac{\mu_{I}}{I} \frac{1}{\left[J(J+1)(2J+1) \right]^{1/2}}\braket{\gamma_{J}J\pi||\boldsymbol{T}^{(1)}||\gamma_{J}J\pi}
\end{equation}
and
\begin{equation}
    B_{J} = 2 Q \left(\frac{J(2J-1)}{(J+1)(2J+1)(2J+3)} \right)^{1/2}\braket{\gamma_{J}J\pi||\boldsymbol{T}^{(2)}||\gamma_{J}J\pi} \ .
\end{equation}
The magnetic dipole moment $\mu_{I}$ results from the nuclear reduced matrix element $\braket{I||\boldsymbol{M}^{(1)}||I}$ while the electric quadrupole moment $Q$ results from $\braket{I||\boldsymbol{M}^{(2)}||I}$. Both nuclear moments are experimentally known quantities \cite{STONE20161}. \\

\section{Canonical and natural orbitals} 
\label{sec:NO}
\subsection{On the use of Natural Orbitals in Quantum Chemistry}

In their pioneer work, Kutzelnigg~{\it et al.}~\cite{Kut:64a,AhlKut:68a} investigated the direct determination of natural orbitals and natural expansion coefficients of many-electron wavefunctions. NOs have been introduced in modern methods and algorithms in Quantum Chemistry for the development of ab initio methods for electron correlation in molecules~\cite{Wer:87a,Knoetal:2000a}. Although they are known  for generating compact CI wave functions of high quality, as discussed by Bytautas~{\it et al.}~\cite{Bytetal:2003a}, the question of their usefulness  for generating an efficient expansion of the wave function is still open and is the subject of recent investigations by Giesbertz~\cite{Gie:2014a}.
In the framework of Density-Functional Theory (DFT), natural (spin)-orbitals are explored to describe the motion of individual electrons in small molecular systems and develop a method for correlated electronic structure calculation~\cite{Gebetal:2016a}.

More in line with the present work, 
Engels~{\it et al.}~\cite{Engetal:96a} observed that improved isotropic hyperfine coupling constants of radicals can be obtained if natural orbitals are used instead of molecular orbitals, thanks to the increased compactness of the wave function.

\subsection{On the use of Natural Orbitals in Atomic Physics}

The use of natural orbitals (and approximate Brueckner orbitals) has been investigated by Lindgren~{\it et al.}~\cite{Linetal:76a} in their study of the hyperfine interaction in alkali atoms in many-body perturbation calculations. They found that when these orbitals are used to evaluate the polarization and lowest-order correlation effects, some important higher-order effects are automatically included. They also observed that for the alkali atoms, the modification of the HF orbitals towards Brueckner or natural orbitals affects considerably the valence orbital, pulling the electron closer to the nucleus, increasing the hyperfine interaction. Brueckner orbitals were later used on Ca$^{+}$ and Ca hyperfine structures by M\aa{}rtenson-Pendril \textit{et al.}~\cite{Maartensson1984} and Salomonson~\cite{Salomonson1984}, following the work of Lindgren~{\it et al.}~\cite{Linetal:76a}.

Natural orbitals have never been used for optimization strategies in multiconfiguration variational calculations of atomic properties to the knowledge of the authors except for studies of states of two-electron systems, where the expansion is also referred to as the ``reduced form''.  The latter, applied to pair-correlation functions, leads to non-orthogonal orbital sets~\cite{Fro:73a,Fro:77a}. \\
\noindent
The electron density and natural orbitals~\cite{Low:55a} of non-relativistic multiconfiguration expansions can be computed using the DENSITY program~\cite{Boretal:2010a} designed for  ({\sc ATSP}2K) program package. The approach of the DENSITY program has been extended to the relativistic framework~\cite{Lietal:2019a}. Introducing, for each $\kappa$, the density matrix 
\begin{equation}
\boldsymbol{\rho}^{\kappa} = \rho^{\kappa}_{n,n'} \ ,
\end{equation}
with elements
\begin{equation}
\rho^{\kappa}_{nn'} = \sum_{ij} c_i \nu^{ij}_{nn'\kappa} c_j \ ,
\end{equation}
where $\nu_{nn' \kappa}^{i,j}$ are angular coefficients, the computation of which are detailed in [28], 
the natural orbitals  $\boldsymbol{\widetilde{\phi}}^{\kappa} =  \widetilde{\phi}_{n \kappa}$ are obtained from the eigenvectors $\boldsymbol{U}^{\kappa}$ of the density matrix \begin{equation}
\label{eq:diag}
    \boldsymbol{U}^{\dagger} \boldsymbol{\rho}\boldsymbol{U}=\boldsymbol{\tilde{\rho}} \; ,
\end{equation}
\begin{equation}
    \label{eq:egvc}
    \boldsymbol{\tilde{\phi}}=\boldsymbol{\phi}\boldsymbol{U}  \; .
\end{equation}
Written explicitly the natural orbitals are
\begin{equation}
{\widetilde P}_{n' \kappa}(r) = \sum_n u_{n,n'}^{\kappa} P_{n \kappa}(r) \ , 
\end{equation}
\begin{equation}
{\widetilde Q}_{n' \kappa}(r) = \sum_n u_{n,n'}^{\kappa} Q_{n \kappa}(r)  \ .
\end{equation}
The eigenvalues of the density matrix can be interpreted as the occupation numbers $\eta^{\kappa}_{n'}$ of the NOs with
\begin{equation}
    0\leq \eta^{\kappa}_{n'} \leq 1 \ , \quad \sum_{\kappa}\sum_{n'}\eta^\kappa_{n'}=N \ .
\end{equation}

\subsection{Rotations in complete active spaces towards reduced forms}
\label{sec:rotCAS}
\subsubsection{Non-uniqueness of the wave functions}
The ASF (\ref{eq:ASF}) is expanded over the CSFs belonging to an active space. The active space is built by defining an active set (AS) of orbitals and generation rules to generate CSFs. If all possible CSFs are generated, then the active space is \textit{complete}~\cite{Froetal:97b}. The size of complete active spaces (CASs) grows rapidly with the number of electrons. Considering the ground state of sodium, an active set of orbitals may include all spectroscopic $\{1s,2s,2p_{-},2p,3s\}$ orbitals and correlation orbitals up to a given principal maximum number, $n$, with a priori all angular momenta $l \leq n-1$. An active set of orbitals is defined as the set of all orbitals from which and to which electron substitutions are allowed. In this work the active set of orbital is denoted $nl$. If the $l$-value is not specified then it corresponds to its maximum value $l_{max}=n-1$. The reference spectroscopic orbitals are often split into core orbitals and valence orbitals. The former are the innermost closed shells, e.g., $\{1s,2s,2p_{-},2p\}$ for sodium,  whilst the latter correspond the outermost orbitals, e.g., $\{3s\}$ for sodium. The generation of all possible CSFs would require to allow up to eleven simultaneous substitutions from the reference orbitals to the AS. The smallest CAS built on the $n=3$ active set already generates over 100~000 CSFs and the CAS $n=4$ generated over 750~000~000 CSFs.

When the active space is complete, then the total wave function supports rotations amongst orbitals of the same $\kappa$-symmetry within the active set, i.e., transforms $\Psi(\Gamma J\pi)$ to itself, with different mixing coefficients. Applying any unitary transformation to the radial part of the orbitals leaves the total wave function invariant, and therefore the total energy or any other observable. The non-uniqueness of the wave function is discussed in greater details in~\cite{Froetal:97b} in the non-relativistic framework. If the active space is complete, the equivalent reduced form of the two-electron pair wave function is often selected to enhance the convergence~\cite{Fro:77a}. The corresponding orbitals are precisely the natural orbitals.
\subsubsection{Reduced form of the MCDHF Ca expansion}
\label{sec:redform}
A small test-case is described for the [Ar]$4s^{2} \ ^{1}\!S_{0}$ calcium ground state to illustrate how the active space is reduced in the NO basis. The [Ar]$4p^{2} \ ^{1}\!S_{0}$ and [Ar]$3d^{2} \ ^{1}\!S_{0}$ configurations are added to the reference to form the multireference set. In a layer-by-layer (LBL) optimisation scheme, the CSF active space is progressively increased by allowing all SD from the $\{3d_{-},3d,4s,4p_{-},4p\}$ valence orbital set to the $7f$ active set, containing seven $s$ orbitals, six $p/p_{-}$ orbitals, five $d/d_{-}$ orbitals and four $f/f_{-}$ orbitals. The active space is complete and therefore supports any transformation, including the one producing the natural orbitals. In Table~\ref{tab:reduced} are reported the mixing coefficients of the first few CSFs of the ASF expansion in both LBL and NO bases. The mixing coefficients of all CSFs of the form $n\kappa n'\kappa$ with $n\neq n'$ approach zero within the numerical accuracy in the natural-orbital basis, bringing the ASF in its reduced form as expected for a single pair-correlation function~\cite{Fro:73a,Fro:77a}. Note that the leading configurations, such as $3d^{2} \ ^{1}\!S_{0}$, $4s^{2} \ ^{1}\!S_{0}$ or $4p^{2} \ ^{1}\!S_{0}$, gain weight by the transformation to the natural orbitals. These results follow closely the work on neutral beryllium of Borgoo {\it et al.}~\cite{Boretal:2010a} in non-relativistic MCHF calculations.
\begin{table}[]
    \centering
    \caption[]{\raggedright \small{The mixing coefficients of the first 20 CSFs of the Ca $4s^{2} \ {}^{1}\!S_{0}$ ground state expansion are given in the LBL and NO orbital bases. Note how in the NO orbital basis the weight is concentrated to CSFs resulting from double excitations to the same orbital.}}
    \label{tab:reduced}
    \begin{tabular}{lccrccr}
    \hline
    \hline
CSFs & & & \multicolumn{1}{c}{LBL} & & & \multicolumn{1}{c}{NO} \\
\hline
    $3d^{2} $  &&&  $-$0.04020926   &&&   $-$0.04021725\\
   $3d_{-}^{2}$   &&&  $-$0.03245460   &&&   $-$0.03246118\\
   $3d4d$   &&&  $-$0.00031844   &&&   $-$0.00000002\\
   $3d5d$   &&&  0.00070691   &&&   $-$0.00000004\\
   $3d6d$  &&&  0.00011109   &&&   $-$0.00000000\\
   $3d7d$   &&&  0.00009347   &&&   $-$0.00000005\\
   $3d_{-}4d_{-}$   &&&  $-$0.00025446   &&&   $-$0.00000003\\
   $3d_{-}5d_{-}$   &&&  0.00057827   &&&   $-$0.00000007\\
   $3d_{-}6d_{-}$  &&&  0.00008856   &&&   0.00000001\\
   $3d_{-}7d_{-}$   &&&  0.00007677   &&&   $-$0.00000010\\
   $4s^{2}$  &&&  0.95883438   &&&   0.95946635\\
   $4s5s$   &&&  $-$0.02923057   &&&   $-$0.00000006\\
    $4s6s$   &&&  $-$0.01969643   &&&   $-$0.00000008\\
    $4s7s$  &&&  $-$0.00071432   &&&   $-$0.00000014\\
    $4p^{2}$   &&&  0.22286640   &&&   0.22290319\\
    $4p_{-}^{2}$   &&&  0.15940321   &&&   0.15942915\\
    $4p5p$  &&&  0.00273900   &&&   $-$0.00000015\\
   $4p6p$  &&&  $-$0.00286431   &&&   0.00000023\\
    $4p7p$   &&&  $-$0.00058552   &&&   0.00000048\\
   $4p_{-}5p_{-}$   &&&  0.00194042   &&&   $-$0.00000004\\
   \multicolumn{1}{c}{$\vdots$} &&& \multicolumn{1}{c}{$\vdots$} &&& \multicolumn{1}{c}{$\vdots$} \\
   $7f_{-}^{2}$   &&&  $-$0.00018046   &&&   $-$0.00018168\\
\hline
\hline
    \end{tabular}
\end{table}

\subsubsection{Superiority of NO in non-equivalent orbital bases}
\label{sec:superiority}
The previous section presented the reduced form of the calcium ground state. We should emphasise here that the wave function is left unchanged by the transformation to the natural orbitals and so is the total binding energy or any observable $\braket{\hat{O}}_{\Psi}$. The two representations (LBL and NO) are therefore equivalent. Strengthened by this experience, another set of rotations may be applied to the LBL basis. These rotations are arbitrary chosen to highly mixed pairs of orbitals of the same symmetry. For each such pairs, $(4s,5s);(6s,7s);(4p_{-},5p_{-});(3d,4d);\cdots$, maximal rotations with coefficients $(\pm 1/\sqrt{2} , 1/\sqrt{2})$ are selected. The result is a strong rearrangement of the radial orbitals in all symmetries. Since the active space is complete, the energy and the wave function remain unchanged by the transformations. These three bases, LBL, NO and ROT are therefore strictly equivalent. They can then be used in more extensive CI calculations including e.g., core-valence and core-core correlation. The VV+CV+CC active space is generated by allowing all SD excitations from the $\{3s,3p_{-},3p,3d_{-},3d,4s,4p_{-},4p\}$ orbitals to the $7f$ active set. The introduction of core and core-valence correlation breaks the completeness of the active space. The extended CSF expansion beyond valence excitations does not support rotations anymore, and the resulting computed energies differ in each basis. As anticipated the complete VV active space shows no energy difference between the orbital bases. When considering the VV+CV+CC active space, the NO orbital basis provides the lowest estimation of the total energy - by 67 cm$^{-1}$ with respect to the original LBL value, while the rotated (ROT) orbital basis gives a disastrous result, far too high energy by around 27~500 cm$^{-1}$. This sequence,  $E_{\text{NO}} < E_{\text{LBL}} < E_{\text{ROT}}$,  indicates that these orbital bases are non-equivalent.
The NO basis leading to the lower total binding energy might be optimal.
Other criteria than the total energy can be used to differentiate the orbital bases. Amongst them, the mean radius of the spectroscopic valence orbital and its associated generalized occupation number provide valuable information. The mean radii, $\braket{r_{4s}}_{\text{LBL}}=4.20207$, $\braket{r_{4s}}_{\text{NO}}=4.13432$ and $\braket{r_{4s}}_{\text{ROT}}=6.33763$, and the corresponding generalized occupation numbers, $\eta^{4s}_{\text{LBL}} = 1.83997$, $\eta^{4s}_{\text{NO}} =1.84115$ and $\eta^{4s}_{\text{ROT}} = 0.883646$, point towards the same direction, i.e., the $4s$ natural orbital is the most contracted one and has its $\eta^{4s}$ closer to the expected $\eta^{4s}=2$ for the $[Ar]4s^{2} \ ^{1}S_{0}$ pure configuration. We could therefore argue that the NOs are better suited to core-valence and core-core correlation, which explains the lower energy value. The larger generalized occupation number enhances the weight of the dominant CSF. Finally, one should notice that if the VV+CV+CC active space was complete, i.e., would allow simultaneously up to $10$ substitutions from the $\{3s,3p_{-},3p,3d_{-},3d,4s,4p_{-},4p\}$ orbitals, the invariance of the total wave function would be preserved and the computed energies identical. In the present case, we started with a complete active space, leading to equivalent orbital bases. This equivalence is then spoiled by the introduction of core excitations that make the expansion incomplete. To illustrate the impact of missing CSFs that prevent the invariance property under orbital rotations, a simpler case is considered in App.~\ref{sec:way} in which an incomplete active space is progressively enriched until completeness is reached.

\section{Hyperfine structure of the sodium ground state}
\label{sec:NaGS}
The $1s^{2}2s^{2}2p^{6}3s \ ^{2}S_{1/2}$ ground state of neutral sodium is investigated using the full relativistic MCDHF method. Two different orbital bases are optimized and compared. They both span the same active space since their active set of orbitals (AS) are identical, i.e., the number of available orbitals per $\kappa$-symmetry is the same.
\subsection{LBL orbital basis}
\label{sec:MCDHF_OB}
 The first orbital basis, labelled LBL, is optimized in three steps following a layer-by-layer scheme:
\begin{enumerate}[(i)]
    \item{The reference orbitals are determined by solving the $N$-electron Dirac-Hartree-Fock (DHF) equations. The $1s$, $2s$, $2p_{-}$ and $2p_{+}$ orbitals are then frozen to their DHF solution.}
    \item{Seven layers of correlation orbitals are optimized based on a core-valence active space generated by allowing SrD substitutions from the DHF configuration. The largest core-valence active set is represented by its maximum principal quantum number $n=9$ and maximum angular quantum number $l$, corresponding to $9h$. The first correlation layer optimizes together all $n=3$ orbitals, including the spectroscopic $3s$ orbital which is therefore no longer solution to the DHF equations.}
    \item{Two additional layers of correlation orbitals are optimized based on a core-valence plus core-core active space generated by allowing all SD substitutions from the DHF configuration. The largest active set is $11h$ and corresponds to the final LBL orbital basis set.}
\end{enumerate}

\subsection{Natural-orbital basis}
The natural-orbital basis requires the prior knowledge of the LBL orbital basis and is computed in two steps as follows:
\begin{enumerate}[(i)]
\item{The density matrix based on the mixing coefficients of the ASF expansion in the LBL representation is evaluated and diagonalized for the $9h$ CV active space according to Eq.~\ref{eq:diag}. Its eigenvectors (\ref{eq:egvc}) provide the required coefficients to build the natural orbitals by linear combinations of the LBL orbitals.}
\item{As in step (iii) for generating the LBL basis, two additional layers of correlation orbitals are optimized based on a CV+CC active space generated by allowing all SD substitutions from the DHF configuration. The orbitals with $1\leq n \leq 9$ are the natural orbitals resulting from the transformation described in the previous step (i) while orbitals with $n \geq 10$ are not transformed to NOs.}
\end{enumerate}
The transformation to the natural orbitals leads to a radial re-organization of the LBL orbitals within each $\kappa-$symmetry. Table~\ref{tab:egvc} shows the vector compositions of each $g^{\text{NO}}_{-}$ orbitals in the LBL basis. The $5g^{\text{NO}}_{-}$, $8g^{\text{NO}}_{-}$ and $9g^{\text{NO}}_{-}$ are all largely dominated (more than 85$\%$) by the original $5g_{-}$, $8g_{-}$ and $9g_{-}$, respectively. The other two $g_{-}$ orbitals are highly mixed and reveal a change in their dominant character ($6g_{-} \rightleftharpoons 7g_{-}$). The principal quantum number of each natural orbital is chosen accordingly to the density matrix eigenvalues sorted in decreasing order, i.e., the dominant component of the eigenvectors does not necessarily define the principal quantum number (as shown for the $6g^{\text{NO}}_{-}$ and $7g^{\text{NO}}_{-}$ orbitals). Moreover, when the active space is symmetric with respect to the $6g_{-}$ and $7g_{-}$ orbital labels, permutations are unimportant. However the strong radial mixing of $n\kappa n'\kappa$ orbital pairs might strongly perturb the representation of the transformed total wave function as shown in App.~\ref{sec:way}.
\begin{table}[!ht]
    \centering
    \caption[]{\raggedright \small{Vector compositions of the natural orbitals in the LBL orbital basis for the $g_{-}$ symmetry.}}
    \label{tab:egvc}
    \begin{tabular}{c|rrrrr}
    \hline
    \hline
       & \multicolumn{1}{c}{$5g_{-}$} & \multicolumn{1}{c}{$6g_{-}$} & \multicolumn{1}{c}{$7g_{-}$} & \multicolumn{1}{c}{$8g_{-}$} & \multicolumn{1}{c}{$9g_{-}$} \\
       \hline
    $5g^{\text{NO}}_{-}$  & $-$0.969 &  0.173  &$-$0.174 &  0.021   & $-$0.005 \\
    $6g^{\text{NO}}_{-}$  &  0.245   &  0.630  &$-$0.715 &  0.165   & $-$0.072 \\
    $7g^{\text{NO}}_{-}$  &  0.018   &  0.757  & 0.633   & $-$0.157 &  0.038   \\
    $8g^{\text{NO}}_{-}$  & $-$0.020 &  0.007  & 0.241   &  0.934   & $-$0.264 \\
    $9g^{\text{NO}}_{-}$  &  0.007   &  0.020  &$-$0.014 &  0.275   &  0.961   \\
    \hline
    \hline
    \end{tabular}
\end{table}
\subsection{Configuration interaction}
Higher order effects are taken into account in CI calculations for both bases independently. The Breit interaction is added to the Dirac-Coulomb Hamiltonian (see Sec.~\ref{sec:MCDHF}) together with QED corrections such as the self-energy correction or the vacuum polarization  correction~\cite{Froetal:2016b}. The active space is further increased to include S, D, triple (T) and quadruple (Q) substitutions from the reference to the largest active set of orbitals, i.e., $11h$. Due to the limits of the available computational resources, triple and quadruple excitations are only allowed to a subset of orbitals. The largest active space therefore combines SD excitations from the reference orbitals to the $11h$ active set with T substitutions to the $6f$ active subset and Q substitutions to the 4 active subset, leading to 617~695 CSFs.
\subsection{Full-variational, layer-by-layer and natural orbitals in MCHF}
\label{sec:N0_MCHF}
For light elements such as sodium, non-relativistic calculations are often used as guidelines for relativistic calculations. Moreover, in the context of the layer-by-layer optimization strategy and natural orbitals, the non-relativistic MCHF method enables us to compare the LBL and full-variational (FV) strategies, contrary to MCDHF for which only the LBL approach is available. The non-relativistic active space is expanded similarly as the relativistic one with only two noticeable differences: the angular quantum number is limited to $l_{max}=6$ instead of $l_{max}=5$ and the density matrices are evaluated for each $l$-symmetry instead of each $\kappa$-symmetry. Two different natural orbital bases can be found for FV and LBL bases. They are labelled FV$_{\text{NO}}$ and LBL$_{\text{NO}}$, respectively.

Table~\ref{tab:ANR} compares the hyperfine constant $A_{1/2}$ values of Na $3s \ ^{2}\!S_{1/2}$ obtained with FV, FV$_{\text{NO}}$, LBL and LBL$_{\text{NO}}$ optimization strategies for an increasing active space. In the FV and LBL approaches, the corresponding natural orbitals (FV$_{\text{NO}}$ and LBL$_{\text{NO}}$, respectively) are computed before including CC correlation and triple substitutions. A relativistic correction is finally included by multiplying the non-relativistic results by a DHF/HF factor, as described in~\cite{Jonetal:96b}. The largest calculations in the FV, FV$_{\text{NO}}$ and LBL$_{\text{NO}}$ show relative differences below $0.5\%$ while the LBL orbital basis leads to relative differences around $1.5\%$. The use of NOs restores the agreement between the FV and LBL bases. These observations comfort us in using the natural-orbital basis in the relativistic LBL optimization strategy.
\begin{table}[!ht]
\begin{center}
\caption[]{\raggedright \small{Hyperfine constant $A_{1/2}$ of the sodium ground state for an increasing active space using the MCHF method. The full-variational (FV) and layer-by-layer (LBL) optimization strategies are compared. In both strategies, the natural orbitals are computed after seven layers of CV correlation orbitals. A DHF/HF correction is included to account for relativistic effects~\cite{Jonetal:96b}}.}
\label{tab:ANR}
\begin{tabular}{lcccccc}
\hline
\hline
&& \multicolumn{5}{c}{$A_{1/2}$(MHz) } \\
\cline{3-7}
 Active set && \multicolumn{1}{c}{FV}&\multicolumn{1}{c}{FV$_{\text{NO}}$}&& \multicolumn{1}{c}{LBL}&\multicolumn{1}{c}{LBL$_{\text{NO}}$} \\
\hline
HF               & & 626.645 & && 626.645 \\
MCHF CV &&& \\
3         & & 683.576 & && 683.576 \\
4       & & 843.205 & && 828.179 \\
5     & & 884.187 & && 859.807 \\
6   & & 907.610 & && 885.089 \\
7 & & 927.596 & && 897.176 \\
$8i$ & & 928.142 & && 928.097 \\
$9i$ & & 927.113 ~&~  925.846 & ~~~~~ & 927.742 ~&~926.409 \\
MCHF CV+CC & \\
$10i$&& 864.982 &878.130 & &842.763&878.142   \\
$11i$&& 865.345  &879.018&&843.267&878.441 \\ 
&&&&\\
CI CV + CC + T &&&& \\
SD[$11i$]&&&&\\
$\cup$ T[4]&&868.465&877.001&&849.721&876.401\\
$\cup$ T[5]&&871.197&876.784&&856.027&875.938\\
$\cup$ T[6]&&870.596&874.483&&857.876&873.025\\
&&&&&\\
$\times$ DHF/HF (1.0137) &&882.523 &886.463 &&869.629&
884.985\\
&&&&&\\
Expt. && \multicolumn{5}{c}{$885.813~064~4(5)$~\cite{Becetal:74a}} \\
\hline
\hline
\end{tabular}
\end{center}
\end{table}

\subsection{Layer-by-layer and natural orbitals in MCDHF}
\label{sec:NaGS_comp}
    Table~\ref{tab:A_comparison} compares the $A_{1/2}$ magnetic dipole hyperfine constant computed in the two different orbital bases. The convergence of the $A_{1/2}-$value is observed along the active space expansion. The difference between the LBL and the NO bases is less than 2 MHz when CV correlation alone is included, i.e., at the end of the second optimization step (see Sec.~\ref{sec:MCDHF_OB}). The introduction of CC correlation surprisingly degrades the agreement between the two bases up to a difference of $\sim36$~MHz. The triple substitutions have opposite influence on the $A_{1/2}$ hyperfine constant, reducing the difference to $\sim16$~MHz. According to Engels~\cite{Engels1087}, the effect of the triple substitutions is to increase the value of the hyperfine constant. The LBL orbital basis exhibits this  behaviour whereas the NO basis shows an unexpected decrease in its $A_{1/2}-$value. \\
\begin{table}[!ht]
\begin{center}
\caption[]{\raggedright \small{Relativistic hyperfine constant $A_{1/2}$ (MHz) along the active space expansion in the LBL and NO orbital bases.}}
\label{tab:A_comparison}
\begin{tabular}{lccc}
\hline
\hline
& \multicolumn{3}{c}{$A_{1/2}$(MHz) } \\
\cline{2-4}
Active set & \multicolumn{1}{c}{LBL} && \multicolumn{1}{c}{NO} \\
\hline
DHF &633.698\\
MCDHF+CI CV & \\
$3$      &691.693 \\
$4$    &837.150 \\
$5$  &870.354 \\
$6h$&895.195 \\
$7h$&906.639 \\
$8h$&939.435 \\
$9h$&938.813 && 937.083 \\
&&\\
MCDHF+CI CV + CC && \\
$10h$ & 852.679 && 888.676   \\
$11h$& 852.806 && 888.725 \\ 
&&\\
CI CV + CC + T && \\
SD[$11h$]&&\\
$\cup$ T[$4$]& 859.307 && 886.605\\
$\cup$ T[$5f$]& 865.388 && 885.925\\
$\cup$ T[$6f$]& 866.826 && 883.113\\
&& \\
CI CV + CC + T + Q && \\
SD[$11h$]&&\\
$\cup$ TQ[$4$] & 962.146 && 889.111 \\
$\cup$ T[$6f$] $\cup$ Q[$4$] & 869.945 && 885.841\\
&&&\\
Expt. & \multicolumn{3}{c}{$885.813~064~4(5)$~\cite{Becetal:74a}} \\
\hline
\hline
\end{tabular}
\end{center}
\end{table}
    The form of the non-relativistic hyperfine operators suggests that $s$-orbitals contribute the most to dipole magnetic constant of a $^{2}\!S$ state~\cite{Aouetal:2018a}. The radial dependency of the relativistic hyperfine operators is $\propto r^{-2}$ and therefore the mean radii of $s$-orbitals provide valuable information to understand the difference between the two orbital bases. Table~\ref{tab:MeanR} presents the mean radii and the expectation value of $r^{-2}$ of the spectroscopic and core-valence correlation $s$-orbitals in the LBL and NO orbital bases. The transformation to natural orbitals results in a contraction of the spectroscopic $3s$ orbital $$\braket{r_{3s}}_{\text{LBL}}=4.12007 \ \to \ \braket{r_{3s}}_{\text{NO}}=4.05088$$ $$\braket{r^{-2}_{3s}}_{\text{LBL}}=0.43488 \ \to \ \braket{r^{-2}_{3s}}_{\text{NO}}=0.48120 \ ,$$ which affects more the $\braket{r^{-2}_{3s}}$ expectation value ($10.7\%$) than its mean radii ($1.7\%$). A strong mixing between the $7s$ and the $8s$ orbitals $$\braket{r_{7s}}_{\text{LBL}}=6.16109 \ \to \ \braket{r_{7s}}_{\text{NO}}=2.12338$$ and $$ \braket{r_{8s}}_{\text{LBL}}=0.60187 \ \to \ \braket{r_{8s}}_{\text{NO}}=3.62967 $$ is also observed. Moreover, the LBL $7s$ orbital exhibits a large mean radius, larger than the valence $3s$ orbital. However, from an optimization based on CV correlation, we would expect its mean radius to lie between the $2s$ core orbital and the $3s$ valence orbital. This diffuse correlation orbital already gives a hint that the layer-by-layer optimization strategy leading to the LBL orbital basis is not well suited for hyperfine structure calculations.
\begin{table}[!ht]
    \centering
    \caption[]{\raggedright \small{Mean radii $\braket{r}$ and $\braket{r^{-2}}$ expectation value of the spectroscopic and core-valence correlation $s$-orbitals in the LBL and NO orbital bases.}}
    \label{tab:MeanR}
    \begin{tabular}{c|ccccccc}
    \hline
    \hline
    Orbital & $\braket{r}_{\text{LBL}}$ & & $\braket{r}_{\text{NO}}$&&$\braket{r^{-2}}_{\text{LBL}}$ & & $\braket{r^{-2}}_{\text{NO}}$ \\
    \hline
    $1s$ & 0.14257 & & 0.14373&&229.695&&228.337\\
    $2s$ & 0.77744 & & 0.77649&&14.6385&&15.9002\\
    $3s$ & 4.12007 & & 4.05068&&0.43488&&0.48120\\
    $4s$ & 1.88585 & & 2.07477&&10.9785&&9.5595\\
    $5s$ & 1.83942 & & 1.89431&&22.1684&&22.0093\\
    $6s$ & 1.71470 & & 1.82421&&41.7204&&46.2714\\
    $7s$ & 6.16109 & & 2.12338&&5.4562&&135.812\\
    $8s$ & 0.60187 & & 3.62967&&346.067&&173.626\\
    $9s$ & 1.11220 & & 1.83795&~~~&168.514&&207.676\\
    \hline
    \hline 
    \end{tabular}
\end{table}{}

\subsection{Quadruple excitations}
 Quadrupole (Q) excitations are known to play a small role into the computations of the hyperfine constant~\cite{Engetal:96a}. Since the corresponding number of CSFs growths rapidly with the active space expansion, they can only be included in a very limited subset of orbitals. They were neglected in original non-relativistic calculations~\cite{Jonetal:96b}. Since then, the available computational resources have increased, allowing us to expand further the active space. The CI calculations including Q excitations are shown in Table~\ref{tab:A_comparison}. Two sets of additional CI are performed. The first one is based on the SD[$11h$] $\cup$ TQ[$4$] active space, with identical active set for the triple and quadruple excitations, and gives $A_{1/2}=889.111$ MHz to be compared with $A_{1/2}=886.605$ obtained with the SD[$11h$] $\cup$ T[$4$] active space expansion in the NO basis. The introduction of quadruple excitations therefore increases the hyperfine constant value by $2.5$ MHz. The second one is based on the SD[$11h$] $\cup$ T[$6f$] $\cup$ Q[$4$] active space, with slightly larger active set for the triple excitations than quadruple excitations, leading to a $A_{1/2}$ value of 885.841 in the NO basis. The comparison of this value with the one obtained based on the SD[$11h$] $\cup$ T[$6f$] attributes to the quadruple excitations an increase of the hyperfine constant value by $2.7$ MHz, confirming that the effect of the quadruples is almost additive to the effect of the triple excitations. Since the triple and quadruple excitations have opposite effects on the magnetic dipole constant, both should be included in the final CI calculations. As for the triple excitations, the effect of the quadruple excitations is opposite in the LBL and NO bases. Due to limitation in the available computational resources, the 885.8 value is kept as our final estimation of the $A_{1/2}$ value of the Na ground state.
\subsection{Discussion and comparison with other methods and experiments}
The results presented in previous sections showed that natural orbitals are an interesting tool to compute hyperfine structure constants of alkali-like systems. The small difference between the LBL and NO basis when $A_{1/2}$ is computed over the CV correlation active space allows to use one or the other for further calculations. The discrepancy arising from the core-core CSFs raises questions about the physical meaning of the NO and the optimal choice of an orbital basis for CI calculations. The particularities of the NOs are not well known for systems with more than two electrons~\cite{Low:55a}. We observe that in the NO basis the spectroscopic orbitals have a larger generalized occupation number and the corresponding DHF CSFs larger mixing coefficients. The mean radii of the orbitals are another comparison point to differentiate the two bases. From Table~\ref{tab:MeanR}, it was already shown that the spectroscopic $3s$ orbital is more contracted in the NO basis. The contraction of the valence spectroscopic orbital results naturally from the core-valence correlation as it is seen in the fully variational approach~\cite{Jonetal:96b}. The influence of the contraction of the spectroscopic $3s$ orbital is investigated in greater details in Sec.~\ref{sec:spectro}.
Table~\ref{tab:hfs_comp} presents the magnetic dipole hyperfine structure constant of the neutral sodium ground state. The results of the present work are compared to other theoretical models~\cite{Jonetal:96b,SalYnn:91a,Liu89,Safetal:98a} and experiments~\cite{Becetal:74a}. Our final $A$-values are provided for the LBL and NO bases. Couple-Cluster and MBPT-based theories are in agreement with the experimental value with relative error below $0.3\%$. Non-relativistic calculations performed by J\"onsson et al.~\cite{Jonetal:96b} are also in agreement with the experimental value ($0.4\%$). The NO basis result is close to both the non-relativistic CI value ($0.4\%$) and the experiment ($0.001\%$) while the LBL basis leads to a relative error of 1.8$\%$ to the experiment. The extraordinary small error of the NO basis result relative to the experiment is impressive, but for a more decisive comparison information about the uncertainties of the final value is needed. Our goal is to prove the efficiency of the natural-orbitals basis rather than to provide the most accurate value of the magnetic dipole constant of neutral sodium.
\begin{table}[!ht]
    \centering
    \caption[]{\raggedright \small{The hyperfine structure constants of sodium are compared to theoretical methods and experiments.}}
    \label{tab:hfs_comp}
    \begin{tabular}{lcccc}
        \hline
        \hline
         && $3s \ ^{2}S_{1/2}$&& \\
         Method && $A_{1/2}$ (MHz) &&Ref.\\
         \hline
         DHF      && 633.7 && This work\\
         LBL    && 869.9 && This work  \\
         NO       && 885.8 && This work \\
         \hline
         CI       && 882.2 && ~\cite{Jonetal:96b}\\
         CCSD     && 883.8 && ~\cite{SalYnn:91a}\\
         SD && 884.5 && ~\cite{Liu89}\\
         SD && 888.1 && ~\cite{Safetal:98a}\\
         &&&& \\
         Expt.    && 885.8$^{\dagger}$ && ~\cite{Becetal:74a}\\
        \hline
        \hline
         \footnotesize{$^{\dagger}$885.813~064~4(5)} \\
    \end{tabular}
\end{table}{}

\subsection{On the crucial role of the spectroscopic $3s$ orbital}
\label{sec:spectro}
\subsubsection{Sub-spaces contributions analysis}
The unexpected variation of the magnetic dipole hyperfine constant when CC correlation is added to the CV active space leads us to investigate in more details the different contributions. The active space based on the $9h$ active set is split into three pieces or \textit{sub-spaces}. The first one corresponds to the valence subspace (V) including only CSFs generated from the single excitations $3s \to ns$ ($4 \leq n \leq 9)$. The second one is the core-valence subspace (CV) including CSFs generated by all single substitutions from the core and restricted double substitutions (at most one excitation in the core). The third one, the core subspace (C), contains all double substitutions from the core. Each of these sub-spaces contributes with different intensities to the final $A_{1/2}$ value. Six contributions are computed and tabulated in Table~\ref{tab:contrib}, namely the V-V, CV-CV and C-C \textit{diagonal} contributions and the V-CV, V-C and CV-C \textit{off-diagonal} contributions.
Their sum, corresponding to the total $A_{1/2}$ value, is also given for each basis and each active space. The CV active space calculations show that the V-V, V-CV and CV-CV contributions provided in both bases are close to each other. In the NO basis, the V-V interaction is slightly stronger than in the LBL basis (by $\approx 2$ MHz). This is consistent with the property of the NO basis to increase the mixing coefficient of the leading DHF CSF. Similar feature is observed in the CV+CC calculations, even though the difference in the V-V interaction is larger ($\approx 29$ MHz). The C-C and V-CV interactions are also larger in the NO basis than in the LBL basis by $\approx 3$ MHz and $\approx 1.7$ MHz, respectively. Independently of the basis, we observe from Table~\ref{tab:contrib} that the effect of CC correlation is mainly indirect since the change in the V-V contribution is more than five times larger than the C-C contribution itself (e.g., for the LBL basis the V-V contribution decreases by $\approx 80$~MHz while the C-C contribution is only $\approx 14$~MHz) as it was already discussed in the analyses of Engels \textit{et al.}~\cite{Engels1087} and Godefroid \textit{et al.}~\cite{Godetal:97b}. The interaction between the valence and core sub-spaces is exactly zero due to the one-body structure of the hyperfine Hamiltonian.
\begin{table}[]
    \centering
    \caption[]{\raggedright \small{Contributions of the V, CV and C sub-spaces to the magnetic dipole hyperfine constant of the ground state of neutral sodium. The effect of CC correlation on $A_{1/2}$ is larger in the LBL basis, leading to a lower $A_{1/2}$-value. Note that its effect is mostly indirect since the decrease in the V-V contribution is much larger than the C-C contribution itself. See text for further discussion.}}
    \label{tab:contrib}
    \begin{tabular}{cccccc}
        \hline
        \hline
        & \multicolumn{5}{c}{$A_{1/2}$ (MHz)} \\
        \cline{2-6}
         & \multicolumn{2}{c}{CV} & & \multicolumn{2}{c}{CV+CC} \\
        \cline{2-3}
        \cline{5-6}
        Interaction & LBL & NO && LBL & NO \\
        \hline
        V-V & 775.908 & 777.888 && 696.790&727.814  \\
        V-CV & 150.192 & 147.442&&136.823&139.695 \\
        V-C & / & / & & 0.0 & 0.0 \\
        CV-CV & 12.714 & 11.754&&9.206&8.891 \\
        CV-C & / & / && $-$1.894&$-$1.844\\
        C-C & / & / &&14.322&16.098 \\
        &&&&\\
        $\sum$ & 938.813&937.083&&855.247&890.654 \\
         \hline
         \hline
    \end{tabular}
\end{table}{}
\subsubsection{Radial functions of the $3s$ orbital: LBL vs. NO}
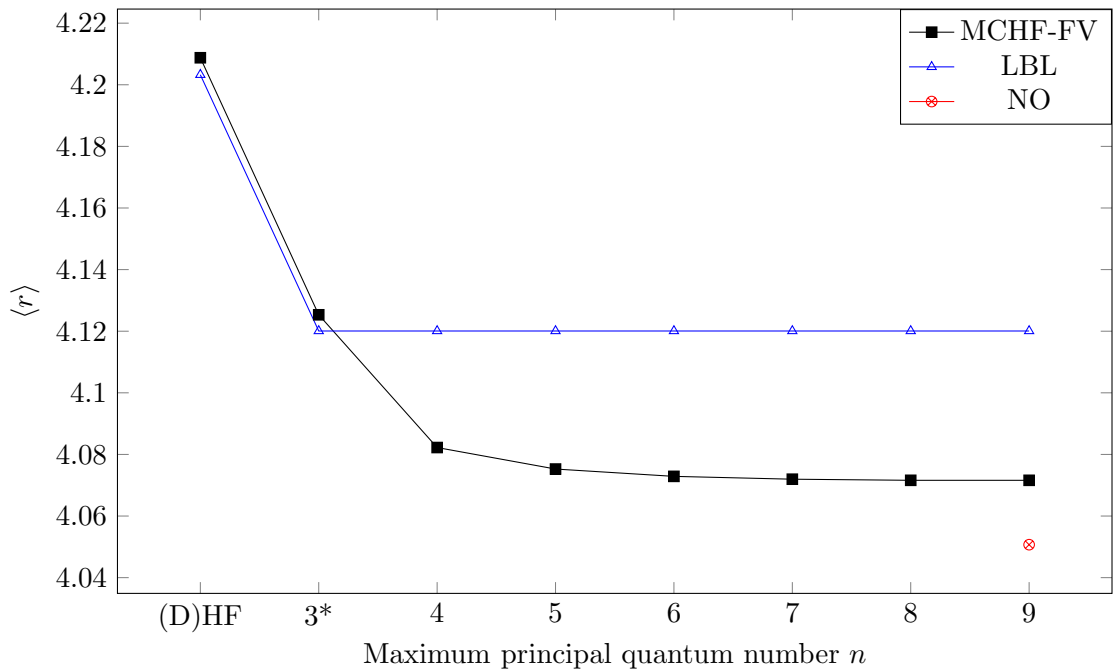
\begin{figure}[!ht]
\begin{center}
\begin{tikzpicture}
\begin{axis}[width=0.9\textwidth,height=0.4\textheight,cycle list name=auto,
xlabel = {Maximum principal quantum number $n$},
legend columns=1,
ylabel = {$\braket{r}$},
xtick={1,2,3,4,5,6,7,8},
xticklabels={$\text{(D)HF}$,3*,4,5,6,7,8,9},
mark size=2.0pt, 
legend entries={\small{MCHF-FV},\small{LBL},\small{NO}},
legend style={at={(1,1)}}]

\addplot+[color=black,solid,mark=square*,mark options={solid}]coordinates{(1 , 4.20877)(2 , 4.12533)(3 , 4.08224)(4 , 4.07526)(5 , 4.07288)(6 , 4.07196)(7 , 4.07160)(8 , 4.07160)}; 

\addplot+[color=blue,solid,mark=triangle]coordinates{(1,4.20325) (2 , 4.12007)(3 , 4.12007)(4 , 4.12007)(5 , 4.12007)(6 , 4.12007)(7 , 4.12007)(8, 4.12007)}; 

\addplot+[color=red,solid,mark=otimes]coordinates{(8 , 4.0507)}; 

\end{axis}
\end{tikzpicture}
\end{center}
\vspace*{-0.8cm}
\caption[]{\raggedright \small{Mean radii of the $3s$ orbital in the LBL basis, NO basis and the non-relativistic FV scheme in which all orbitals are optimized together. The layer star next to the $n=3$ value is there as a reminder that the spectroscopic $3s$ orbitals was varied along with the correlation $3p$ and $3d$ orbitals.}}
\label{fig:3sr}
\end{figure}

 Sec.~\ref{sec:NaGS_comp} showed the difference in mean radii for the $s$-orbitals between the LBL and NO orbital bases. Table~\ref{tab:MeanR} put in evidence the diffuse $7s$ orbital in the LBL basis as well as a more contracted spectroscopic $3s$ orbital in the NO basis. The radial integral of the electronic magnetic dipole operator of the diagonal DHF CSF is $-0.026857$ in the LBL basis and $-0.030312$ in the NO basis. The small difference between the $3s$ and $3s_{\text{NO}}$ orbitals is large enough to create a spectacular $\sim 11 \%$ difference in the hyperfine radial expectation value. Fig.~\ref{fig:Pdiff}  displays the difference of the absolute value the large components $P(r)$ of the spectroscopic $3s$ orbital in both bases. The $3s$ natural orbital is slightly more contracted than the LBL one as attests the three "bumbs" close to the nucleus, which result from a larger transformed $|P_{3s}^{\text{NO}}(r)|$ than the original $|P_{3s}(r)|$ when $r \lessapprox 4$. This leads to a higher value of its radial hyperfine integral in the calculations of the $\braket{1s^{2}2s^{2}2p^{6}3s \ {}^{2}\!S_{1/2} || \boldsymbol{T}^{(1)}||1s^{2}2s^{2}2p^{6}3s \ {}^{2}\!S_{1/2}}$ reduced matrix element.
%
%
\definecolor{mycolor1}{rgb}{0.00000,0.44700,0.74100}%
\begin{figure}
\begin{tikzpicture}

\begin{axis}[%
width=0.9\textwidth,
height=0.4\textwidth,
scale only axis,
xmin=0,
xmax=20,
xlabel style={font=\color{white!15!black}},
xlabel={r},
ymin=-0.015,
ymax=0.01,
ylabel style={font=\color{white!15!black}},
ylabel={$\text{$|$P}_{\text{3s}}\text{(r)$|$ $-$ $|$P}_{\text{3s}}^{\text{NO}}\text{(r)$|$}$},
axis background/.style={fill=white},
legend style={legend cell align=left, align=left, draw=white!15!black}
]
\addplot [color=mycolor1, line width=1.0pt]
  table[row sep=crcr]{%
0	0\\
0.0001872648166	-3.21416358000001e-05\\
0.0001969686313	-3.37983140999999e-05\\
0.0002071699712	-3.55392571999999e-05\\
0.0002178943449	-3.73687316000001e-05\\
0.0002291685691	-3.92912172999999e-05\\
0.000241020835	-4.13114186000001e-05\\
0.0002534807797	-4.34342752e-05\\
0.0002665795594	-4.56649735000001e-05\\
0.0002803499278	-4.80089584e-05\\
0.0002948263182	-5.04719464e-05\\
0.000310044929	-5.30599383e-05\\
0.0003260438146	-5.57792333e-05\\
0.0003428629806	-5.86364429e-05\\
0.0003605444838	-6.16385066999998e-05\\
0.0003791325369	-6.47927069999999e-05\\
0.0003986736199	-6.81066839999999e-05\\
0.0004192165957	-7.15884589999999e-05\\
0.0004408128324	-7.52464429999999e-05\\
0.0004635163318	-7.90894639999999e-05\\
0.0004873838645	-8.31267780000001e-05\\
0.0005124751118	-8.7368098e-05\\
0.0005388528149	-9.1823607e-05\\
0.0005665829316	-9.65039840000001e-05\\
0.0005957348019	-0.000101420428\\
0.0006263813206	-0.000106584678\\
0.0006585991198	-0.000112009039\\
0.000692468761	-0.000117706407\\
0.0007280749357	-0.000123690297\\
0.0007655066781	-0.000129974867\\
0.000804857587	-0.000136574952\\
0.0008462260601	-0.000143506085\\
0.0008897155402	-0.000150784535\\
0.0009354347736	-0.000158427335\\
0.0009834980822	-0.00016645231\\
0.001034025649	-0.000174878117\\
0.00108714382	-0.000183724275\\
0.001142985418	-0.000193011201\\
0.001201690076	-0.000202760242\\
0.001263404586	-0.000212993716\\
0.001328283266	-0.00022373495\\
0.001396488348	-0.000235008308\\
0.001468190378	-0.000246839239999999\\
0.001543568651	-0.000259254317\\
0.00162281165	-0.000272281266\\
0.001706117525	-0.000285949014000001\\
0.001793694583	-0.000300287723000001\\
0.001885761813	-0.000315328831\\
0.001982549431	-0.000331105091\\
0.002084299456	-0.0003476506\\
0.002191266317	-0.00036500085\\
0.002303717486	-0.000383192747\\
0.002421934149	-0.000402264654\\
0.00254621191	-0.000422256419\\
0.002676861529	-0.000443209401000001\\
0.002814209696	-0.000465166495000001\\
0.002958599855	-0.000488172154999999\\
0.003110393056	-0.000512272407999999\\
0.00326996886	-0.000537514865\\
0.003437726291	-0.000563948724\\
0.003614084829	-0.000591624767999999\\
0.003799485463	-0.000620595359\\
0.00399439179	-0.00065091439\\
0.004199291179	-0.000682637310000001\\
0.004414695984	-0.00071582104\\
0.004641144829	-0.000750523899999999\\
0.004879203955	-0.000786805560000001\\
0.005129468634	-0.00082472697\\
0.005392564657	-0.000864350240000001\\
0.005669149901	-0.00090573846\\
0.005959915974	-0.000948955629999998\\
0.006265589943	-0.0009940664\\
0.006586936151	-0.00104113592\\
0.006924758131	-0.00109022956\\
0.007279900615	-0.00114141267\\
0.007653251643	-0.00119475022\\
0.008045744788	-0.00125030646\\
0.008458361486	-0.00130814451\\
0.008892133495	-0.00136832592\\
0.009348145471	-0.00143091013\\
0.009827537681	-0.00149595392\\
0.01033150885	-0.00156351072\\
0.01086131918	-0.00163362995\\
0.01141829347	-0.00170635622\\
0.01200382443	-0.00178172841\\
0.01261937622	-0.00185977877\\
0.01326648801	-0.00194053184\\
0.01394677794	-0.00202400324\\
0.01466194708	-0.00211019847\\
0.01541378372	-0.00219911153\\
0.01620416786	-0.00229072333999999\\
0.01703507585	-0.00238500017000001\\
0.01790858541	-0.00248189189\\
0.01882688076	-0.00258133002\\
0.01979225812	-0.00268322574\\
0.02080713143	-0.00278746766\\
0.02187403842	-0.00289391955\\
0.02299564689	-0.0030024178\\
0.02417476146	-0.00311276888\\
0.02541433053	-0.0032247465\\
0.02671745366	-0.00333808877\\
0.02808738934	-0.00345249519\\
0.02952756313	-0.00356762349\\
0.03104157621	-0.00368308641\\
0.0326332144	-0.00379844847\\
0.03430645762	-0.00391322257\\
0.03606548986	-0.00402686669999999\\
0.0379147096	-0.00413878062\\
0.03985874088	-0.00424830258\\
0.04190244476	-0.00435470639\\
0.04405093159	-0.00445719835\\
0.04630957369	-0.00455491481\\
0.04868401885	-0.00464691985\\
0.05118020441	-0.00473220362\\
0.05380437215	-0.00480968109999999\\
0.05656308384	-0.00487819168\\
0.05946323771	-0.00493649956\\
0.06251208564	-0.00498329515\\
0.06571725135	-0.00501719762000001\\
0.06908674942	-0.00503675887999999\\
0.07262900535	-0.00504046899999999\\
0.07635287662	-0.0050267635\\
0.08026767486	-0.00499403247999999\\
0.0843831891	-0.00494063208000001\\
0.08870971026	-0.00486489825\\
0.09325805691	-0.00476516327\\
0.09803960227	-0.00463977503\\
0.1030663027	-0.00448711942\\
0.1083507276	-0.00430564603\\
0.1139060907	-0.00409389707\\
0.1197462834	-0.00385053996\\
0.1258859092	-0.0035744032\\
0.1323403203	-0.00326451585\\
0.1391256562	-0.00292015023\\
0.1462588837	-0.00254086749\\
0.1537578395	-0.00212656597999999\\
0.1616412751	-0.00167753144\\
0.169928903	-0.00119448837\\
0.1786414467	-0.00067865171\\
0.1878006921	-0.00013177761\\
0.197429542	-0.000443788354\\
0.2075520736	-0.00104506686\\
0.2181935986	-0.00166840663\\
0.2293807261	-0.0023094541\\
0.24114143	-0.00296313375\\
0.253505118	-0.00362364101\\
0.2665027059	-0.00428445084000001\\
0.2801666944	-0.00493834397000001\\
0.2945312505	-0.00557745356\\
0.3096322932	-0.00619333404\\
0.3255075828	-0.00677705369999999\\
0.342196816	-0.00731931180000001\\
0.3597417245	-0.00781058009999999\\
0.3781861797	-0.0082412671\\
0.3975763023	-0.00860190319999998\\
0.4179605777	-0.00888334220000001\\
0.4393899773	-0.00907697439999999\\
0.4619180858	-0.00917494629999999\\
0.485601235	-0.00917037870000001\\
0.5104986453	-0.0090575789\\
0.536672573	-0.00883223979999997\\
0.5641884668	-0.00849162170000001\\
0.5931151306	-0.00803471149999999\\
0.6235248962	-0.00746235820000002\\
0.6554938037	-0.00677738010000001\\
0.6891017923	-0.00598464230000001\\
0.7244328992	-0.00509109759999998\\
0.7615754708	-0.00410578260000001\\
0.8006223827	-0.0030397665\\
0.8416712725	-0.00190603439000001\\
0.884824784	-0.000719301589999999\\
0.9301908233	0.000504245200000003\\
0.9778828292	0.00174728091\\
1.028020057	-0.000831309419\\
1.080727874	-0.00421923385\\
1.13613808	-0.00541154597\\
1.194389228	-0.00655102965\\
1.255626975	-0.0076210201\\
1.32000445	-0.00860614899999998\\
1.387682627	-0.00949257140000001\\
1.45883074	-0.0102680999\\
1.533626694	-0.0109222525\\
1.612257518	-0.0114462345\\
1.694919831	-0.011832882\\
1.781820332	-0.0120765957\\
1.873176316	-0.0121732936\\
1.969216222	-0.0121204073\\
2.0701802	-0.0119169351\\
2.176320711	-0.0115635593\\
2.287903162	-0.0110628199\\
2.405206568	-0.010419331\\
2.528524249	-0.00964001470000003\\
2.658164562	-0.00873431990000001\\
2.794451676	-0.00771439559999998\\
2.937726379	-0.00659517970000001\\
3.088346934	-0.00539436979999997\\
3.24668997	-0.0041322587\\
3.413151427	-0.00283141220000005\\
3.588147545	-0.00151618990000002\\
3.772115906	-0.000212119700000013\\
3.965516527	0.00105485319999998\\
4.16883301	0.00225919680000003\\
4.382573751	0.00337667650000001\\
4.607273215	0.00438523159999998\\
4.843493267	0.00526580160000001\\
5.091824579	0.00600305709999999\\
5.35288811	0.00658599300000001\\
5.627336655	0.00700835129999999\\
5.915856478	0.00726884620000001\\
6.219169028	0.00737117819999999\\
6.538032745	0.00732382469999998\\
6.873244954	0.00713961770000002\\
7.225643861	0.00683512010000001\\
7.596110647	0.00642983\\
7.98557167	0.0059452518\\
8.395000788	0.0054038867\\
8.825421785	0.0048281971\\
9.277910938	0.00423961124\\
9.753599707	0.00365762794000001\\
10.25367756	0.00309907801\\
10.77939495	0.00257758768\\
11.33206645	0.00210327241\\
11.91307403	0.00168267005\\
12.5238705	0.00131890258\\
13.16598317	0.00101203463\\
13.84101766	0.00075958414\\
14.55066192	0.000557126639\\
15.29669041	0.000398932828\\
16.0809686	0.000278582674\\
16.90545759	0.000189507951\\
17.77221904	0.000125429479\\
18.6834203	8.06712718e-05\\
19.64133984	5.03501758e-05\\
};
\addlegendentry{LBL - NO}

\end{axis}
\end{tikzpicture}%
\caption[]{\raggedright \small{Difference between the absolute value of the large component radial function of the spectroscopic $3s$ orbital in the LBL and in NO basis.}}
\label{fig:Pdiff}
\end{figure}
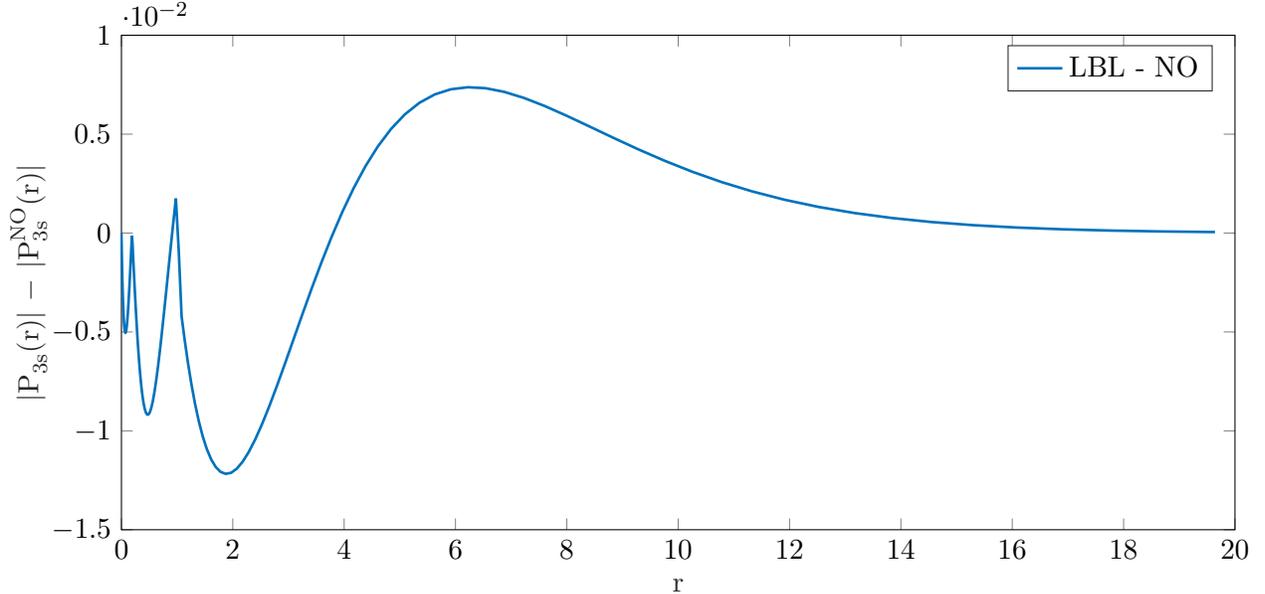
The contraction of the spectroscopic orbital is a direct effect of core-valence correlation. This becomes obvious when the non-relativistic FV optimization strategy of J\"onsson et al.~\cite{Jonetal:96b} (c.f.~\ref{sec:NR_hfs}) is analyzed. We reproduced their computations by allowing all correlation orbitals to vary together with the valence $3s$ orbital for which the mean radius was computed after each layer. In Fig.~\ref{fig:3sr} is plotted the mean radius of the $3s$ orbital along the expansion of the active space characterized by its maximum principal quantum number. For comparison, the LBL and natural $3s$ orbitals mean radii are also shown on the figure. The black line, corresponding to the non-relativistic calculations, clearly shows that the $3s$ orbital is progressively contracted as the active space is expanded, i.e., the core-valence correlation has the direct effect of contracting the valence orbital. The LBL $3s$ orbital is frozen after the optimization of the $n=3$ layer. The LBL optimization strategy is therefore lacking of variational freedom which is partially recovered when computing the natural-orbital basis. Indeed the radial re-organization caused by the diagonalization of the density matrix mixes correlation orbitals and spectroscopic orbitals leading to a more contracted $3s$ orbital.\\

\subsubsection{Artificial contractions}
According to the previous section, the $3s$ orbital is not contracted enough in the LBL approach. The introduction of correlation should compensate for a too diffuse DHF $3s$ orbital. However it is frozen in a layer-by-layer approach. To confirm the strong link between the contraction of the $3s$ orbital (through its mean radius) and the large hyperfine structure constant value, two artificial ways of contracting the $3s$ orbital are explored:\\
\begin{enumerate}[(i)]
\item{Mixing with diffuse orbital \\
The first layer, corresponding to the $n=3$ active space, optimizes together the $3s,3p-,3p,3d-,3d$ orbitals. An additional (artificial) hydrogenic $6s$ orbital is added to the set of optimized orbitals so that the interaction between the diffuse $6s$ orbital and the spectroscopic $3s$ orbital results in a contraction of the latter. Only the $n=3$ orbitals are kept, the $6s$ orbital is removed and the optimization of layers $n=4,5,6,7,8,9$ is performed as described in Sec.~\ref{sec:NaGS}. The corresponding orbital basis is labelled LBL-$6s$.}\\
\item{Extented optimal level optimization (EOL) \\
The first layer, corresponding to the $n=3$ active space, optimizes together the $3s,3p-,3p,3d-,3d$ orbitals by minimizing simultaneously the two $J=1/2$ lowest levels. Relative weights between the lowest and second lowest levels are chosen to be (20,1) or (15,1) in two separate calculations. Such weight partitions affect only slightly the results obtained by optimizing on the lowest level only. These two sets of calculations are labelled LBL-W$20$ and LBL-W$15$, respectively.}
\end{enumerate}

The $A_{1/2}$ hyperfine constants computed with the different optimization strategies are reported in Table~\ref{tab:A3s2} along with the active space expansion. Since these strategies were not built on rational arguments - except the need for a more contracted $3s$ orbital - computations do not include the two core-core correlation layers nor the triple excitations in CI. A single CI calculation is performed for each strategy to include CC correlation which is already enough to appreciate the difference between the LBL and NO bases. The mean radii of the $3s$ orbital are also given in Table~\ref{tab:A3s2}. A high correlation is observed between the change in the mean radius and the hyperfine structure constants. Indeed, going from left to right, the mean radius of the $3s$ orbital decreases, corresponding to a contraction of the orbital, while the $A_{1/2}$ CI CV+CC value increases. It is remarkable that the NO provides similar values for all four bases in both the CV and CV+CC active spaces.
\begin{table}[!ht]
\begin{center}
\caption[]{\raggedright \small{Hyperfine constant $A_{1/2}$ for different optimization processes (see text).}}
\label{tab:A3s2}
\begin{tabular}{lcccc}
\hline
\hline
& \multicolumn{4}{c}{$A_{1/2}$(MHz) } \\
\cline{2-5}
Active set & \multicolumn{1}{c}{LBL} & \multicolumn{1}{c}{LBL-$6s$} & \multicolumn{1}{c}{LBL-W$20$}& \multicolumn{1}{c}{LBL-W$15$} \\
\hline
DHF &633.698&633.698 &633.698&633.698\\
MCDHF+CI CV &&& \\
$3$          & 691.693 & 692.023 & 723.437 & 736.966  \\
$4$        & 837.150 & 844.857 & 856.808 & 862.500  \\
$5$      & 870.354 & 879.427 & 887.737 & 895.054  \\
$6$    & 895.195 & 903.684 & 915.213 & 920.594  \\
$7h$    & 906.639 & 912.244 & 935.625 & 941.586  \\
$8h$    & 939.435 & 938.593 & 938.594 & 941.357  \\
$9h$    & 938.813 & 938.034 & 938.041 & 938.617  \\
$9h$-NO & 937.083 & 936.320 & 936.394 & 936.833  \\
&&&& \\
CI CV+CC: &&&& \\
$9h$     & 855.247 & 860.768 & 876.257 & 883.897 \\
$9h$-NO  & 890.654 & 889.932 & 890.486 & 890.390 \\
&&&& \\
&\multicolumn{4}{c}{$\braket{r}$} \\
$3s$ & 4.12007 & 4.10243 & 4.06620 & 4.04693 \\
$3s$-NO & 4.05068 & 4.05484 & 4.04769 & 4.04951\\
\hline
\hline
\end{tabular}
\end{center}
\end{table}

\subsection{Assessing the reliability of NO \ldots}
Rotations in incomplete active spaces perturb the wave function, sometimes leading to disastrous results as  shown in Sec.~\ref{sec:superiority} and App.~\ref{sec:way}. Since the negative impact of rotations in incomplete active spaces has been established, it is important to assess the reliability of the NO basis. The small example presented in App.~\ref{sec:way} already shows that the NO corresponds to rotations with small angles and therefore only slightly perturb the wave function.

In sodium, due to the single valence electron, core-valence correlation had to be included from the very beginning. The $2p3s \to nln'l'$ and $2s3s \to nln'l'$ classes of substitutions generate an incomplete active space. The opening of the $n=2$ shell would require the simultaneous excitations of nine electrons (two from the $2s$, six from the $2p$ and one from the $3s$) to recover completeness. Fortunately, the LBL orbital basis, as presented in Sec.~\ref{sec:MCDHF_OB}, is not so far from the natural-orbital basis, i.e., the corresponding rotations have small angles. Indeed, all reference orbitals are only slightly perturbed as shown by the leading contribution of their vector composition (at least 99.995 $\%$ in their analogous orbital in the LBL basis). Nevertheless, it is important to assess the quality of the natural-orbital basis, proving that the error due to the NO rotations in our specific incomplete active space remains small.

\subsubsection{\ldots through overlaps}
Since the wave function is not invariant, the relation
\newcommand{\vardbtilde}[1]{\tilde{\raisebox{0pt}[0.85\height]{$\tilde{#1}$}}}
\begin{equation}
\label{eq:Gamma}
    \ket{\Psi\{\phi\}} =  \ket{\tilde{\Psi}\{\tilde{\phi}\}} + \ket{\tilde{\Gamma}\{\tilde{\phi}\}}
\end{equation}
defines $\ket{\tilde{\Gamma}\{\tilde{\phi}\}}$ as the gap between the wave function $|\Psi\{\phi\} \rangle$ built on the LBL orbital basis $\{\phi\}$ and the transformed wave function 
$|\widetilde{\Psi}\{\widetilde{\phi}\}\rangle$ built on the NO orbital basis $\{\widetilde{\phi}\}$. The 
projection of $\ket{\Psi\{\phi\}}$ on itself,
\begin{equation*}
    \underbrace{\braket{\Psi\{\phi\}|\Psi\{\phi\}}}_{=1} =  \braket{\Psi\{\phi\}|\tilde{\Psi}\{\tilde{\phi}\}} + \underbrace{\braket{\Psi\{\phi\}|\tilde{\Gamma}\{\tilde{\phi}\}}}_{:=\gamma_{1}} 
\end{equation*}
allows to introduce
\begin{equation}
\label{eq:gamma1}
    \gamma_{1} = 1 -  \braket{\Psi\{\phi\}|\tilde{\Psi}\{\tilde{\phi}\}} \ ,
\end{equation}
as a measure of the gap, as a function of the overlap between the LBL and transformed wave functions. Since the two orbital bases are nonorthogonal, the biorthogonalization method~\cite{Olsetal:95a} is employed to evaluate the overlap
$\braket{\Psi\{\phi\}|\tilde{\Psi}\{\tilde{\phi}\}}$. The biorthogonalization provides two additional bases, $\{\phi'\}$ and $\{\tilde{\phi}'\}$ such that $\braket{\phi'_{i}|\tilde{\phi}'_{j}} = \delta_{ij}$. The counter-transformation~\cite{Olsetal:95a} acts on the mixing coefficients so that the total wave function is invariant. If $c'_{i}$ and $\tilde{c}'_{i}$ are the mixing coefficients resulting from the counter-transformations for the LBL and NO wave functions respectively, then
\begin{equation}
\label{eq:wfnBIO}
\left \{
\begin{aligned}
\ket{\Psi\{\phi\}} &= \sum_{i=1}^{N}c_{i}\ket{\Phi_{i}\{\phi\}} = \sum_{i=1}^{N}c'_{i}\ket{\Phi'_{i}\{\phi'\}} \\
\ket{\tilde{\Psi}\{\tilde{\phi}\}} &= \sum_{i=1}^{N}\tilde{c}_{i}\ket{\tilde{\Phi}_{i}\{\tilde{\phi}\}} = \sum_{i=1}^{N}\tilde{c}'_{i}\ket{\tilde{\Phi}'_{i}\{\tilde{\phi}'\}}  \ .
\end{aligned}
\right .
\end{equation}
Starting from Eq.~\ref{eq:gamma1} and using the relations of Eq.~\ref{eq:wfnBIO} together with the biorthogonality conditions, $\gamma_{1}$ resumes to
\begin{equation}
    \gamma_{1} = 1 - \sum_{i=1}^{N}c'_{i}\tilde{c}'_{i} \ .
\end{equation}
The overlap represents a measure of the error made by transforming the LBL basis to the NO basis. It is therefore necessary to ensure that atomic properties converge with respect to the NO transformation. The natural orbitals are the orbitals that diagonalize the density matrix. If the active space was complete, re-computing the natural orbitals would be trivial since the density matrix would already be diagonal by definition. If the active space is incomplete, the density matrix re-calculated in the NO basis is no longer diagonal. The diagonalization of this new density matrix provides another NO basis. It is crucial to show that an iterative transformation to NO converges to a diagonal density matrix. Applying iteratively Eq.~\ref{eq:Gamma}, the wave function is written
\begin{equation*}
    \ket{\Psi\{\phi\}} =  \ket{\vardbtilde{\Psi}\{\vardbtilde{\phi}\}} + \ket{\vardbtilde{\Gamma}\{\vardbtilde{\phi}\}} + \ket{\tilde{\Gamma}\{\tilde{\phi}\}} \ .
\end{equation*}
The "second-order" overlap is therefore related to both $\gamma_{1}$ and $\gamma_{2}$ by the relation
\begin{equation*}
   \gamma_{2} + \gamma_{1} = 1- \braket{\Psi\{\phi\}|\vardbtilde{\Psi}\{\vardbtilde{\phi}\}} 
\end{equation*}
which is easily generalized for a number $x$ of transformations as
\begin{equation*}
    \braket{\Psi\{\phi\}|\Psi^{(x)}\{\phi^{(x)}\}} = 1 - \sum_{j=1}^{x} \gamma_{x} \ .
\end{equation*}

\subsubsection{\ldots through iterative transformations}
Iterative transformations to the natural orbitals are applied to the sodium ground state to assess the validity of the results presented in Sec.~\ref{sec:NaGS_comp}. Each iteration consists in the following two-step procedure:
\begin{enumerate}[(i)]
    \item{For a given orbital basis, $\{\phi^{(i)}\}$ with $0 \leq i \leq x-1$ where $x$ is the total number of iterations, the ASF expansion coefficients are determined through CI calculation.}
    \item{A density matrix is computed using the mixing coefficients of step (i) and diagonalized. Its eigenvectors provide the rotations coefficients to build the new orbital basis, $\{\phi^{(i+1)}\}$, as linear combinations of $\{\phi^{(i)}\}$ with $0 \leq i \leq x-1$.}
\end{enumerate}
Table~\ref{tab:xconv} presents the energy, the hyperfine constant and the overlap with the LBL wave function for an increasing number of transformations to the NOs. Each quantity is computed for two different active spaces, one based on CV correlation alone and another based on CV+CC correlation. These three properties seem to converge fast with the number of iterations, $x$. This result is reassuring and leads to the conclusion that performing small rotations in an incomplete active space does not destroy the wave function.
\begin{table}[!ht]
\centering
\caption[]{\raggedright \small{Energy and hyperfine constant as a function of the number of times the NOs are computed. The effect of core-core correlation is also shown.}}
\label{tab:xconv}
\begin{tabular}{cccccccccc}
\hline
\hline
&&\multicolumn{2}{c}{Energy} && \multicolumn{2}{c}{$A_{1/2}$}&& \multicolumn{2}{c}{$|\braket{\Psi|\Psi^{(x)}}-1|$}\\ 
\cline{3-4}
\cline{6-7}
\cline{9-10}
$x$ && CV & CV+CC && CV & CV+CC && CV & CV+CC\\
\hline
0 &&$-$162.0480429 & $-$162.4106111 & & 938.813 & 855.247 & & 8.88[$-$16] & 4.33[$-$15] \\
1 &&$-$162.0480373 & $-$162.4105571 & & 937.083 & 890.654 & & 3.82[$-$7] & 5.51[$-$5] \\ 
2 &&$-$162.0480354 & $-$162.4105655 & & 937.523 & 890.140 & & 1.66[$-$7] & 5.32[$-$5]  \\
3 &&$-$162.0480356 & $-$162.4105632 & & 937.391 & 890.284 & & 2.22[$-$7] & 5.38[$-$5]  \\
4 &&$-$162.0480356 & $-$162.4105639 & & 937.429 & 890.242 & & 2.05[$-$7] & 5.36[$-$5]  \\
5 &&$-$162.0480356 & $-$162.4105637 & & 937.418 & 890.254 & & 2.10[$-$7] & 5.36[$-$5]  \\
6 &&$-$162.0480356 & $-$162.4105637 & & 937.421 & 890.250 & & 2.08[$-$7] & 5.36[$-$5]  \\
7 &&$-$162.0480356 & $-$162.4105637 & & 937.420 & 890.252 & & 2.09[$-$7] & 5.36[$-$5]  \\
8 &&$-$162.0480356 & $-$162.4105637 & & 937.421 & 890.251 & & 2.09[$-$7] & 5.36[$-$5]  \\
9 &&$-$162.0480356 & $-$162.4105637 & & 937.420 & 890.251 & & 2.09[$-$7] & 5.36[$-$5]  \\
10&&$-$162.0480356 & $-$162.4105637 & & 937.420 & 890.251 & & 2.09[$-$7] & 5.36[$-$5]  \\
\hline
\hline
\end{tabular}
\end{table}

\section{Sodium-like ions and sodium excited states}
\label{sec:iso}
The unexpected difference between the two orbital bases in the calculations of the neutral sodium hyperfine structure constants leads us to consider  the lightest sodium like-ions $^{25}$Mg$^{+}$, $^{27}$Al$^{2+}$, $^{29}$Si$^{3+}$, $^{31}$P$^{4+}$, $^{33}$S$^{5+}$ and $^{35}$Cl$^{6+}$. For each ion the ground state hyperfine structure constant is computed and monitored along the active space expansion using the exact same optimization strategy as for the sodium ground state. Table~\ref{tab:isoNa} presents the magnetic dipole hyperfine constants of the sodium isoelectronic sequence from Na to Cl along the active space expansion. The observations are similar than for sodium. Firstly, the LBL and NO bases give close results when core-valence correlation alone is included. Secondly, the two bases lead to larger differences when core-core correlation is added through the optimization of two more layers. Finally, the opposite effects of the triple excitations in the LBL and NO bases is confirmed in all considered ions, i.e., the hyperfine structure constant increases in the LBL basis and decreases in the NO basis. The quadruple substitutions were omitted in the calculations within the iso-electronic sequence and the excited states since the goal of these calculations is to observe a similar discrepancy between the LBL and NO bases as for the neutral sodium ground state rather than providing the most accurate values for the hyperfine structure constants. Table~\ref{tab:isoNa} also presents the relative error on the final A-value computed with both bases on the largest active space. Since neither the NO or the LBL basis is \textit{a priori} better, the relative errors are computed as $\frac{2|\text{A}_{\text{NO}} - \text{A}_{\text{LBL}}|}{|\text{A}_{\text{NO}} + \text{A}_{\text{LBL}}|}\times 100\%$, where the reference value is taken as the mean of the two bases. We observe that the difference between the two bases drops below $0.5\%$ already for the Mg$^{+}$ ion. It continues to drop as heavier ions are considered. It reaches its lowest value for the heavier ion, Cl$^{6+}$, for which the relative difference between the two bases is $0.01\%$. The analyse of the sodium ground state presented in Sec.~\ref{sec:NaGS} showed that the $3s$ orbital was more contracted in the NO basis. Going from neutral Na to Cl$^{6+}$, the spectroscopic $3s$ orbital is more contracted due to the stronger attractive potential with the nucleus leading to larger $A$ value. The contraction of the $3s$ orbital due to the interaction with the core however does not depend equally strong on $Z$ explaining the relatively small importance for the more highly charged ions.

\begin{sidewaystable}
    \centering
    \caption[]{\raggedright \small{Magnetic dipole hyperfine constant of the [Ne]$3s \ ^{2}S_{1/2}$ ground state for Na-like ions. The $A_{1/2}$ is given for $^{23}$Na I ($I=3/2$, $\mu=2.2176556\mu_{N}$), $^{25}$Mg II ($I=5/2$, $\mu=-0.85545\mu_{N}$), $^{27}$Al III ($I=5/2$, $\mu=3.6415069\mu_{N}$), $^{29}$Si IV ($I=1/2$, $\mu=-0.555290\mu_{N}$), $^{31}$P V ($I=1/2$, $\mu=1.13160\mu_{N}$), $^{33}$S VI ($I=3/2$, $\mu=0.64382120\mu_{N}$) and $^{35}$Cl VII ($I=3/2$, $\mu=0.8218743\mu_{N}$).}}
    \label{tab:isoNa}
    \begin{tabular}{lcccccccccccccccccccc}
    \hline
    \hline
     & \multicolumn{20}{c}{$A_{1/2}$ (MHz)} \\
    \hline
    & \multicolumn{2}{c}{Na} && \multicolumn{2}{c}{Mg$^{+}$}& & \multicolumn{2}{c}{Al$^{2+}$} && \multicolumn{2}{c}{Si$^{3+}$} && \multicolumn{2}{c}{P$^{4+}$} && \multicolumn{2}{c}{S$^{5+}$} && \multicolumn{2}{c}{Cl$^{6+}$}  \\
    \cline{2-3}
    \cline{5-6}
    \cline{8-9}
    \cline{11-12}
    \cline{14-15}
    \cline{17-18}
    \cline{20-21}
    Active set & LBL & NO && LBL & NO && LBL & NO && LBL & NO && LBL & NO && LBL & NO && LBL & NO\\
    \hline
DHF & 633.698 & && $-$471.811 & && 4055.30 & && $-$5205.17 & && 16168.9 & && 4385.82 & && 7656.53 & \\ 
MCDHF+CI CV & & & \\
$3$ & 691.693 & && $-$494.936 & && 4187.307 & && $-$5329.86 & && 16472.4 & && 4453.19 & && 7755.91 & \\ 
$4$ & 837.150 & && $-$576.279 & && 4743.16 & && $-$5920.79 & && 18038.2 & && 4823.46 & && 8328.85 & \\ 
$5$ & 870.354 & && $-$587.583 & && 4807.10 & && $-$5986.86 & && 18222.6 & && 4870.93 & && 8409.41 & \\ 
$6$ & 895.195 & && $-$607.470 & && 4961.12 & && $-$6155.59 & && 18667.5 & && 4973.66 & && 8562.93 & \\ 
$7h$ & 906.639 & && $-$612.707 & && 4979.66 & && $-$6169.12 & && 18693.5 & && 4978.23 & && 8568.27 & \\ 
$8h$ & 939.435 & && $-$617.675 & && 5001.17 & && $-$6180.33 & && 18712.6 & && 4981.22 & && 8571.36 & \\ 
$9h$ & 938.813 & 937.083 && $-$618.338 & $-$617.636 && 4999.64 & 4995.93 && $-$6186.04 & $-$6182.69 && 18722.2 & 18714.3 && 4982.21 & 4980.48 && 8570.97 & 8568.42 \\ 
MCDHF+CI CV+CC & & & \\
$10h$ & 852.679 & 888.676 && $-$587.105 & $-$596.202 && 4838.59& 4873.52 && $-$6042.21 & $-$6067.10 && 18401.6 & 18446.8 && 4917.36 & 4924.95 && 8484.12 & 8492.66 \\ 
$11h$ & 852.806 & 888.725 && $-$587.499 & $-$596.461 && 4841.14 & 4875.21 && $-$6044.18 & $-$6068.60 && 18407.4 & 18451.6 && 4918.63 & 4926.01 && 8485.81 & 8494.08 \\ 
CI CV+CC+T & & & \\
SD[$11h$]  & & & \\
$\bigcup$ T[$4$] & 859.307 & 886.605 && $-$590.192 & $-$596.185 && 4855.31 & 4875.95 && $-$6056.19 & $-$6070.46 && 18433.2 & 18457.7 && 4923.65 & 4927.53 && 8492.31 & 8496.44 \\ 
$\bigcup$ T[$5f$] & 865.388 & 885.925 && $-$591.889 & $-$596.007 && 4862.32 & 4875.77 && $-$6061.28 & $-$6070.65 && 18442.9 & 18458.9 && 4925.42 & 4927.93 && 8494.48 & 8497.13 \\ 
$\bigcup$ T[$6f$] & 866.826 & 883.113 && $-$592.535 & $-$595.262 && 4866.08 & 4873.46 && $-$6064.63 & $-$6069.40 && 18450.4 & 18457.6 && 4926.88 & 4927.91 && 8496.39 & 8497.36 \\ 
$\frac{2|\text{A}_{\text{NO}} - \text{A}_{\text{LBL}}|}{|\text{A}_{\text{NO}} + \text{A}_{\text{LBL}}|}\times 100\%$& \multicolumn{2}{c}{$1.86\%$} && \multicolumn{2}{c}{$0.45\%$} && \multicolumn{2}{c}{$0.15\%$}&& \multicolumn{2}{c}{$0.08\%$}&& \multicolumn{2}{c}{$0.04\%$} && \multicolumn{2}{c}{$0.02\%$}&& \multicolumn{2}{c}{$0.01\%$}  \\
~\cite{Safetal:98a}& \multicolumn{2}{c}{$888.1$} && \multicolumn{2}{c}{$-597.6$}&& \multicolumn{2}{c}{$4885$}&& \multicolumn{2}{c}{$-6060$}&& \multicolumn{2}{c}{$18407$}&& \multicolumn{2}{c}{$4910$} \\
Expt. & \multicolumn{2}{c}{$885.813$~\cite{Becetal:74a}} && \multicolumn{2}{c}{$-596.254$~\cite{Xuetal:2017a}} \\
\hline
\hline
    \end{tabular}
    \end{sidewaystable}

The $1s^{2}2s^{2}2p^{6}3p \ ^{2}P^\circ_{1/2}$ and $1s^{2}2s^{2}2p^{6}3p \ ^{2}P^\circ_{3/2}$ excited states of sodium are investigated to check if the observations made for the ground state can be generalized. The active space expansion and the optimization of the orbital basis closely follow the strategy employed for the $1s^{2}2s^{2}2p^{6}3s \ ^{2}S_{1/2}$ ground state. Small differences arise since the $3s$ orbital is now a correlation orbital while the $3p$ orbital is spectroscopic. The $J=1/2$ and $J=3/2$ levels are optimized together following an extended optimal level scheme (EOL)~\cite{Grant}. The natural-orbital basis is also computed after seven CV correlation layers. Its computation somewhat differs from the sodium ground state case since the corresponding density matrix is evaluated as the weighted average of the density matrix of each level. The  $A_{1/2}$ and $A_{3/2}$ hyperfine magnetic dipole constants converge to 90.908 MHz and 18.083 MHz, respectively, in the LBL orbital basis and to 94.31 MHz and 18.803 MHz in the natural-orbital basis. The corresponding experimental values are 94.42(19) MHz~\cite{Caretal:92a} and 18.69(6) MHz~\cite{Itano1981}, leading to relative errors on  $A_{1/2}$ of $3.7\%$ and $0.1\%$ for the LBL and natural-orbital bases, respectively and on  $A_{3/2}$ of $3.2\%$ and $0.6\%$, respectively. These results confirm the potential of the natural orbital basis in the calculations of hyperfine structures as already demonstrated for the sodium ground state. They allow to achieve an accuracy below $1\%$ compared to the experiments which is more than a factor of 3 better LBL basis results. The EFG=$B_{3/2}/Q$ values converge to 25.675 MHz/b and 26.599 MHz/b in the LBL and NO bases, respectively. The convergence of $A_{1/2}$, $A_{3/2}$ and EFG along the active space expansion is illustrated in Table~\ref{tab:2P_A1/2}.

\begin{table}[!ht]
\centering
\caption[]{\raggedright \small{Relativistic magnetic dipole hyperfine constants $A_{1/2}$ (MHz) and $A_{3/2}$ (MHz) of the $^{2}\!P_{1/2}$ and $^{2}\!P_{3/2}$ sodium excited states, respectively, along the active space expansion in two different orbital bases. Simiar results are given for the electric field gradient EFG=$B_{3/2}/$Q (MHz/b) of the $J=3/2$ level.}}
\label{tab:2P_A1/2}
\begin{tabular}{lcccccccc}
\hline
\hline
& \multicolumn{2}{c}{$A_{1/2}$(MHz) } & & \multicolumn{2}{c}{$A_{3/2}$(MHz) } &&\multicolumn{2}{c}{$B_{3/2}/Q$(MHz/b) }\\
\cline{2-3}
\cline{5-6}
\cline{8-9}
Active set & \multicolumn{1}{c}{LBL} & \multicolumn{1}{c}{NO}&&\multicolumn{1}{c}{LBL} & \multicolumn{1}{c}{NO} && \multicolumn{1}{c}{LBL} & \multicolumn{1}{c}{NO} \\
\hline
DHF    &64.157&         && 12.744 &        &&15.939  \\
MCDHF+CI CV & &  \\
$3$   &69.348 &         && 12.054 &        &&16.605 \\
$4$   &91.212 &         && 20.611 &        &&27.648 \\
$5f$  &93.643 &         && 21.077 &        &&27.291\\
$6f$  &98.294 &         && 20.300 &        &&27.822\\
$7f$  &100.751&         && 20.770 &        &&28.520\\
$8f$  &101.791&         && 20.544 &        &&28.562\\
$9f$  &101.861& 101.817 && 20.554 & 20.543 &&28.453 & 28.433\\
&&\\
MCDHF+CI CV + CC && \\
$10f$ &87.740 & 94.655  && 17.184 & 18.592     &&24.424 & 26.274\\
$11f$ &87.701 & 94.603  &~~~~& 17.197 & 18.606 &~~~~&24.469 & 26.322\\
&&\\
CI CV + CC + T && \\
SD[$11f$]&&\\
$\cup$ T[$4$] & 88.596 & 94.668 && 17.514 & 18.755 &&24.838 & 26.474 \\
$\cup$ T[$5f$]& 90.073 & 94.594 && 17.943 & 18.874 &&25.395 & 26.616\\
$\cup$ T[$6f$]& 90.908 & 94.316 && 18.083 & 18.803 &&25.675 & 26.599\\
&& \\
Others & \\
CI~\cite{Jonetal:96b} & \multicolumn{2}{c}{94.04}& &\multicolumn{2}{c}{18.80}&&\multicolumn{2}{c}{25.79} \\
SD~\cite{Safetal:98a} & \multicolumn{2}{c}{94.99} &&\multicolumn{2}{c}{18.84}&&\multicolumn{2}{c}{26.85}\\
SD~\cite{Liu89} & \multicolumn{2}{c}{92.4}&&\multicolumn{2}{c}{19.3}&& \\
CCSD~\cite{SalYnn:91a} & \multicolumn{2}{c}{93.02}&&\multicolumn{2}{c}{18.318}&&\multicolumn{2}{c}{26.14} \\
Expt.~\cite{Caretal:92b,Voltz1996} & \multicolumn{2}{c}{94.42(19)}& \multicolumn{2}{c}{18.79(12)}&&\\
\hline
\hline
\end{tabular}
\end{table}

\clearpage
\section{Conclusion}
We report hyperfine structure calculations of the neutral sodium ground state and first excited states. We presented a detailed analysis on the use of the natural orbitals in multiconfiguration methods for hyperfine structure constants and the limitations of the traditional layer-by-layer optimization scheme. Extensive testing was performed on the natural orbital basis to assess its reliability in incomplete active spaces, looking at the basis transformation as any other rotations. The lack of variational freedom of the layer-by-layer scheme was found to be directly related to the shape of the spectroscopic orbitals, inducing a large discrepancy between our computed values and the experimental measurements. The natural-orbital basis method allows us to relax the frozen condition on the spectroscopic orbitals and to modify them according to the correlation model, leading to results in better agreement with experiments and other theoretical works. We provide evidence that natural orbitals have interesting properties and can be considered as a promising alternative to the LBL optimization scheme in more complex systems, e.g., with more than one valence electron. Beside their own particular properties that make the natural orbitals an interesting alternative, the simplicity of their computations could help in assessing the stability of computational atomic properties, i.e., they could participate in the estimation of the theoretical uncertainty.

\begin{acknowledgments}
\noindent 
SS is a FRIA grantee of the Fonds de la Recherche Scientifique$-$FNRS. CFF acknowledges support from the Canada NSERC Discovery Grant 2017-03851. MG acknowledges support from the FWO \& FNRS Excellence of Science Programme (EOS-O022818F). PJ and JE acknowledge support from the Swedish Research Council under contract 2015-04842. Computational resources have been provided by the Shared ICT Services Centre, Universit\'e libre de Bruxelles and by the Consortium des \'Equipements de Calcul Intensif (C\'ECI), funded by the Fonds de la Recherche Scientifique de Belgique (F.R.S.-FNRS) under Grant No. 2.5020.11 and by the Walloon Region.
\end{acknowledgments}
\clearpage
\appendix
\section{The way to completeness: a sodium example}
\label{sec:way}
A small test on the ground state of sodium $1s^{2}2s^{2}2p^{6}3s \ ^{2}\!S_{1/2}$ is used to demonstrate the need for the completeness of the active space to support rotations of the orbital basis. The active space is built very simply by allowing excitations from the $2p_{1/2}$ and $2p_{3/2}$ orbitals to the $3p_{1/2}$ and $3p_{3/2}$ orbitals. The number of simultaneous substitutions is progressively increased from one to six. The corresponding active space is labelled by the number of allowed excitations and built as follows
\begin{equation*}
\begin{aligned}
(0)\qquad&1s^{2}2s^{2}2p^{6}3s \ ^{2}\!S_{1/2} \\
(1)\qquad& + \quad 1s^{2}2s^{2}2p^{5}3s3p \ ^{2}\!S_{1/2}\\
(2)\qquad& + \quad1s^{2}2s^{2}2p^{4}3s3p^{2} \ ^{2}\!S_{1/2}\\
(3)\qquad& + \quad1s^{2}2s^{2}2p^{3}3s3p^{3} \ ^{2}\!S_{1/2}\\
(4)\qquad& + \quad1s^{2}2s^{2}2p^{2}3s3p^{4} \ ^{2}\!S_{1/2}\\
(5)\qquad& + \quad1s^{2}2s^{2}2p3s3p^{5} \ ^{2}\!S_{1/2}\\
(6)\qquad& + \quad1s^{2}2s^{2}3s3p^{6} \ ^{2}\!S_{1/2}\\
\end{aligned}
\end{equation*}
Starting from the reference configuration, the above configurations are progressively added such that the active space (6) becomes a CAS. For each active space ($1\to 6$), the orbital basis set is such that the $\{1s, 2s, 2p_{1/2},2p_{3/2}, 3s\}$ orbitals are solutions to the DHF equations while the orbitals $\{3p_{1/2},3p_{3/2}\}$ are solutions to the corresponding multiconfiguration equations. The $p_{1/2}$ and $p_{3/2}$ orbitals are then rotated according to
\begin{equation}
\label{eq:rot}
\left \{
\begin{aligned}
2p'_{\kappa} &= \frac{1}{\sqrt{2}}2p_{\kappa} + \frac{1}{\sqrt{2}}3p_{\kappa} \\
3p'_{\kappa} &= -\frac{1}{\sqrt{2}}2p_{\kappa} + \frac{1}{\sqrt{2}}3p_{\kappa}
\end{aligned}
\right.
\end{equation}
for $\kappa=1, -2$, inducing a large mixing. For each active space, the energies before and after rotations are computed as well as their difference $\Delta E$. They are displayed in Table~\ref{tab:IAS_p}.
\begin{table}[!h]
\centering
\caption[]{\raggedright \small{Energy differences induced by rotations within the $p$-symmetries either with a 50-50 mixing or a transformation to natural orbitals, as a function of an increased active space.}}
\label{tab:IAS_p}
\begin{tabular}{ccccc}
\hline
\hline
Active space && $\Delta E$ (cm$^{-1}$)&& $\Delta E_{\text{NO}}$ (cm$^{-1}$) \\ 
\hline
1&&612~036&&0.00 \\
2&&995~555&&0.20 \\
3&&441~332&&0.02 \\
4&&134~728&&0.00 \\
5&&18~731 &&0.00 \\
6&&0      &&0.00\\
\hline
\hline
\end{tabular}
\end{table}
Although expected, going from the active space $5$ to the active space $6$ is quite spectacular. Indeed the extra configuration $1s^{2}2s^{2}3s3p^{6} \ ^{2}\!S_{1}$ generates only one CSF. This CSF alone allows to recover the invariance of the energy and the wave function. Its mixing coefficient is $-0.000289$ ($\sim 0.000008 \%$) in the LBL basis and $-0.082971$ ($\sim 0.65 \%$) in the rotated basis. A similar analysis is performed when evaluating the total binding energy in the natural orbital basis. For each active space, the natural orbitals are computed, inducing a mixing within the $p$-symmetries. The corresponding rotations have smaller angles than the 50-50 example, leading energy differences $\Delta E_{\text{NO}}$ close to zero. They are displayed in Table~\ref{tab:IAS_p} beside $\Delta E$.
\newpage
\strut

\end{document}